\documentclass[useAMS,usenatbib]{mn2e}
\def\sp{{\it{Spitzer}}}

\newcommand{\bjdutc}{\ensuremath{\rm {BJD_{UTC}}}}

\newcommand{\bmjdobs} {\rm {BMJD\_OBS}}

\usepackage[pdftex]{graphicx}
\usepackage{amsmath}
\usepackage{amssymb}
\usepackage{deluxetable}
\usepackage[breaklinks,colorlinks,urlcolor=blue,citecolor=blue,linkcolor=blue]{hyperref} 

\title[K and clouds in the atmosphere of
WASP-31b]{\textbf{\textit{HST}} hot-Jupiter
  transmission spectral survey:  detection of potassium in WASP-31b
  along with a cloud deck and Rayleigh scattering}

\author[D. K. Sing et al.]
{D. K. Sing$^{1}$\thanks{E-mail: sing@astro.ex.ac.uk}, 
H. R. Wakeford$^{1}$, 
A. P. Showman$^{2}$,
N. Nikolov$^{1}$,  
J. J. Fortney$^{3}$,\newauthor
A. S. Burrows$^{4}$, 
G. E. Ballester$^{2}$, 
D. Deming$^{5}$, 
S. Aigrain$^{6}$,
J.-M. D\'esert$^{7}$,\newauthor
N. P. Gibson$^{8}$, 
G. W. Henry$^{9}$,
H. Knutson$^{10}$,
A. Lecavelier des Etangs$^{11}$, 
F. Pont$^{1}$, \newauthor
A. Vidal-Madjar$^{11}$,
M. W. Williamson$^{9}$,
P. A. Wilson$^{1}$  
\\
$^{1}$Astrophysics Group, School of Physics, University of Exeter, Stocker Road, Exeter, EX4 4QL, UK\\ 
$^{2}$Lunar and Planetary Laboratory, University of Arizona, Tucson, Arizona 85721, USA\\
$^{3}$Department of Astronomy and Astrophysics, University of California, Santa Cruz, CA 95064, USA\\
$^{4}$Department of Astrophysical Sciences, Peyton Hall, Princeton University, Princeton, NJ 08544, USA\\ 
$^{5}$Department of Astronomy, University of Maryland, College Park, MD 20742 USA\\
$^{6}$Department of Physics, University of Oxford, Denys Wilkinson Building, Keble Road, Oxford OX1 3RH, UK\\ 
$^{7}$CASA, Department of Astrophysical and Planetary Sciences, University of Colorado, 389-UCB, Boulder, CO 80309, USA\\
$^{8}$European Southern Observatory, Karl-Schwarzschild-Str. 2, D-85748 Garching bei Munchen, Germany\\ 
$^{9}$Tennessee State University, 3500 John A. Merritt Blvd., PO Box 9501, Nashville, TN 37209, USA\\ 
$^{10}$Division of Geological and Planetary Sciences, California Institute of Technology, Pasadena, CA 91125 USA\\ 
$^{11}$CNRS, Institut dAstrophysique de Paris, UMR 7095, 98bis boulevard Arago, 75014 Paris, France
}
\begin{document}
\date{Accepted 2014 October 28.  Received 2014 September 4; in original form 2014 July 24}
\pagerange{\pageref{firstpage}--\pageref{lastpage}} \pubyear{2014}
\maketitle

\label{firstpage}

\begin{abstract}
We present {\it Hubble Space Telescope} optical and near-IR transmission spectra
of the transiting hot-Jupiter WASP-31b. 
The spectrum covers 0.3--1.7 $\mu$m at a resolution $R\sim$70, which we combine with {\it Spitzer} photometry to cover the full-optical to IR.  
The spectrum is dominated by a cloud-deck with a flat transmission spectrum 
which is apparent at wavelengths $>0.52\mu$m.  The cloud deck is present at high
altitudes and low pressures, as it covers the majority of the
expected optical Na line and near-IR H$_2$O features.
While Na~{\small I} absorption is not clearly identified, the resulting spectrum
does show a very strong potassium feature detected at the 4.2-$\sigma$
confidence level. 
Broadened alkali wings are not detected, indicating pressures below $\sim$10 mbar.
The lack of Na and strong K is the first indication of a 
sub-solar Na/K abundance ratio in a planetary atmosphere (ln[Na/K]$=-3.3\pm2.8$), which could potentially be explained by Na condensation
on the planet's night side, or primordial abundance variations. 
A strong Rayleigh scattering signature is
detected at short wavelengths, with a 4-$\sigma$ significant slope.  
Two distinct aerosol size populations can explain the 
spectra, with a smaller sub-micron size grain population  
reaching high altitudes producing a blue Rayleigh scattering signature
on top of a larger, lower-lying population responsible for the flat cloud deck
at longer wavelengths.  
We estimate that the atmospheric circulation is sufficiently strong to
mix micron size particles upward to the required 1--10
mbar pressures, necessary to explain the cloud deck.
These results further confirm the importance
of clouds in hot-Jupiters, which can potentially dominate
the overall spectra and may alter the abundances of key gaseous
species.  
\end{abstract}


\section{Introduction}
Hot Jupiters are gas giant planets found in tight orbits around their
parent stars, with a large number now known to transit their host
star and are amenable to atmospheric investigations.  This wide class of
exoplanets span a very large range  
in equilibrium temperatures, with most planets ranging from
1000 to 3000 K depending on the albedo and efficiency of heat transfer
from the day to night side.  At these temperatures, theoretical models
of hot Jupiters have predominately predicted atmospheres which are dominated by water opacity in the
near-IR, with the optical containing alkali features and scattering by
hydrogen molecules (e.g.
\citealt{2000ApJ...537..916S, %seager
2003ApJ...588.1121S, % Sudarsky
2010ApJ...719..341B, %Burrows
2010ApJ...709.1396F}).  %Fortney

 For highly irradiated hot Jupiter atmospheres, the expectation of
 alkali metals leads to very low predicted albedos.  This expectation
 has generally been backed up by observations, both from the detection
 of Na (\citealt{2002ApJ...568..377C,%  Charbonneau et al. 2002; 
2011ApJ...743..203J, %Jensen
2014MNRAS.437...46N, %Nikolov
2012MNRAS.422.2477H, %Huitson
2008ApJ...673L..87R,%Redfield et al. 2008;
2008ApJ...686..658S, 2012MNRAS.426.1663S, %Sing
2012MNRAS.426.2483Z}) % Zhou
and K \citep{ 2011A&A...527A..73S}  %Sing
atoms as well as the generally low albedos found in many
hot Jupiters such as HD~209458b 
\citep{2008ApJ...689.1345R}.  %Rowe

The atmospheric temperatures of the hot Jupiters are close to the
condensation temperature of several abundant components, including
silicates and iron, making clouds and hazes a natural outcome of
chemistry in much the same way as abundant H$_2$O, CO, and CH$_4$
molecules in gaseous form.  The possible
presence of such condensation clouds was considered early on 
\citep{2000ApJ...537..916S,2001ApJ...560..413H}.  %Seager \& Hubbard
According to models, condensates would weaken
spectral features, or mask some of them, depending on the height of
the cloud deck 
\citep{1999ApJ...513..879M, %Marley
2003ApJ...588.1121S, %Sudarsky
2005MNRAS.364..649F}.  %Fortney
There is a growing body of evidence for substantial clouds and hazes
in hot Jupiters, including the transmission spectra of HD~189733b, 
WASP-12b, HAT-P-32b, and WASP-6b \citep{2013MNRAS.432.2917P,
  2011MNRAS.416.1443S, 2013MNRAS.436.2956S, 2013MNRAS.436.2974G, NikolovWasp6}.
% Pont, Sing, Sing, Gibson, Nikolov

Here we present results for WASP-31b from a large {\it Hubble Space Telescope}
({\it HST}) programme ({\it HST} GO-12473; P.I. Sing), consisting of an eight planet optical atmospheric survey of transiting
hot Jupiters.   The overall programme goals are to detect atmospheric features across a
wide range of hot-Jupiter atmospheres enabling comparative
exoplanetology, detect stratosphere causing agents like TiO
\citep{2003ApJ...594.1011H, 2008ApJ...678.1419F}, %Hubbeny, Fortney
and detect alkali atoms as well as hazes and clouds.
In the results reported so far for this survey,
\cite{2013MNRAS.434.3252H} %Huitson
detected H$_2$O in WASP-19b and found TiO was unlikely.
Water was also found in HAT-P-1b \citep{2013MNRAS.435.3481W} 
with the broad-band spectrum also showing Na~{\small I} and a significant
optical absorber but no K~{\small I} absorption \citep{2014MNRAS.437...46N}.
Finally, TiO was ruled out in the very hot planet WASP-12b and the atmosphere was found to
be dominated by aerosols, with corundum a possible candidate that is
particularly relevant given its high condensation temperature
applicable to this very-hot Jupiter \citep{2013MNRAS.436.2956S}. %Sing 

WASP-31b was discovered by \cite{2011A&A...531A..60A} %Andeson
as part of the Wide Angle Search for Planets survey.  With a low mass
(0.46 M$_{Jup}$) and large inflated radius (0.155 R$_{Jup}$) the planet is of particular interest to
atmospheric studies as it has a very low density ($\rho_{pl}$=0.129
$\rho_{Jup}$) and correspondingly low surface gravity of 4.56 m
$s^{-1}$ making the planet very favorable to transmission spectroscopy.
The exoplanet orbits at 0.047 AU around a F6V 6300~K low metallicity
([Fe/H]=-0.19) low activity ~3-5 Gyr star, and has a nominal (zero
albedo, full recirculation) equilibrium temperature of 1570~K \citep{2011A&A...531A..60A}. %Anderson
In this paper, we present new {\it HST} transit observations with the 
Space Telescope Imaging Spectrograph (STIS) and 
Wide Field Camera 3 (WFC3) in spacial scanning mode, combining them with {\it Spitzer}
photometry to construct a high signal-to-noise (S/N) near-UV to infrared
transmission spectrum of WASP-31b, capable of detecting and scrutinising atmospheric constituents.
We describe our observations in Sect. 2, present the analysis of the
transit light curves in Sect. 3, discuss the results in Sect. 4 and
conclude in Sect. 5.

%%%%%%%%%%%%%%%%%%%%%%%%%%%%%%%%%%%%%%%%%%%%%%%%%%%%%%%%%%%%

\section{Observations}

\subsection{\emph{Hubble Space Telescope} STIS spectroscopy}
The overall observational strategy for WASP-31b is nearly identical for each of the eight targets
in the large {\it HST} programme, including the WASP-19b, 
HAT-P-1b, and WASP-12b which have been published in \cite{2013MNRAS.434.3252H,  2013MNRAS.435.3481W, 2014MNRAS.437...46N} and \cite{2013MNRAS.436.2956S}.  %Huitson, Wakeford, Nikolov, Sing

We observed two transits of WASP-31b with the {\it HST} STIS $G$430$L$
grating during 2012 June 13 and 2012 June 26, as well as one transit
with the STIS $G$750$L$ during 2012 July 10.  
The $G$430$L$ and $G$750$L$ data sets 
each contain 43 spectra, which span five spacecraft orbits.
 The $G$430$L$ grating covers
the wavelength range from 2,900 to 5,700~{\AA}, with a resolution $R$ of
$\lambda$/$\Delta\lambda=$530--1,040 ($\sim$2 pixels; 5.5~{\AA}).  
The $G$750$L$ grating covers
the wavelength range from 5,240 to 10,270 ~{\AA}, with a $R=$530--1,040 ($\sim$2 pixels; 9.8~{\AA}).  
Both the STIS data were taken with a wide 52 arcsec$\times2$ arcsec slit to minimize slit light
losses.
The visits of {\it HST} were scheduled such that the third and fourth spacecraft orbits
contain the transit, providing good
coverage between second and third contact, as well as an out-of-transit
baseline time series before and after the transit. 
Exposure times of 279 s were used in conjunction with a 128-pixel wide
sub-array, which reduces the readout time between exposures
to 21 s, providing a 93 per cent overall duty cycle.

The data set was pipeline-reduced with the latest version of {\small CALSTIS}, and 
cleaned for cosmic ray detections with a customised procedure based on median-combining difference images
following \cite{2014MNRAS.437...46N}.  The $G$750$L$ data set was defringed
using contemporaneous fringe flats.  
The mid-time of each exposure was converted into BJD$_{\rm TBD}$ for use
in the transit light curves \citep{2010PASP..122..935E}. %Eastman

\begin{figure*}
{\centering
  \includegraphics[width=1\textwidth,angle=0]{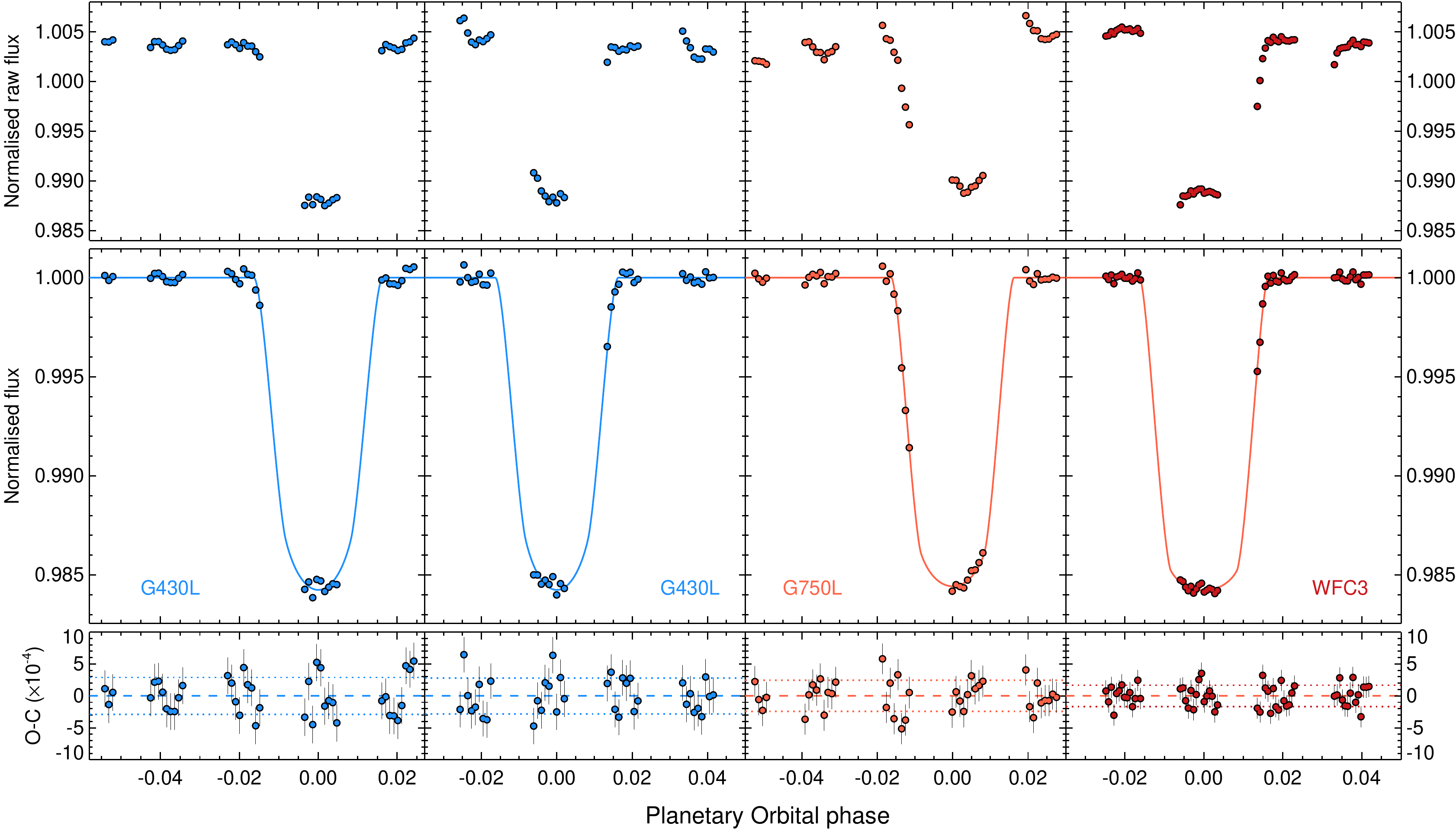}}
\caption[]{{\it HST} STIS normalized white-light curves for the three
  WASP-31b transits taken with the STIS $G$430$L$, STIS $G$750$L$, and WFC3.
Top panels: Raw light curves normalized to the mean out-of-transit raw flux. 
Middle panels: Detrended light curves along with the best-fitting
limb-darkened transit model superimposed with continuous lines.
Lower panels: Observed minus modelled light curve residuals, compared
to a null (dashed lines) and 1$-\sigma$ level (dotted lines).}
\label{figwhite}
\end{figure*}

The spectral aperture extraction was done with {\small IRAF} using a
13-pixel-wide aperture with no background subtraction, which minimises the out-of-transit standard deviation of the white-light curves.
The extracted spectra were then Doppler-corrected to a common rest frame through cross-correlation, 
which helped remove sub-pixel wavelength shifts in the dispersion direction. 
The STIS spectra were then used to create both a white-light photometric 
time series (see Fig. \ref{figwhite}), and custom wavelength bands covering the spectra, 
integrating the appropriate wavelength flux from each exposure for
different bandpasses.  

Like the other targets in our {\it HST} programme, the STIS light curves of
WASP-31 show instrument-related systematic effects primarily due to the well known
thermal breathing of {\it HST}, which warms and cools the telescope during the 96 min 
day/night orbital cycle, causing the focus to vary.  The first
exposure of each spacecraft orbit has also consistently
been found to show significantly lower fluxes than the remaining exposures.
Similar to our previous STIS analysis \citep{2013MNRAS.436.2956S,
  2013MNRAS.434.3252H, 2014MNRAS.437...46N}, we intended to discard the entire
first orbit to minimise the effect of the breathing trend, but found
for two of the three {\it HST} visits, a few of the
exposures taken toward the end of the first orbit could be used in the
analysis and fit with the same systematics model as used for the subsequent
orbits.  
In addition, we set the exposure time of the first image of each spacecraft orbit to be 1
second in duration, such that it could be discarded without
significant loss in observing time.  We find that the second exposures taken
for all five spacecraft orbits (each 279 s in duration) do not
in general show the first-exposure systematic trend, with an exception of the third
orbit of the $G$750$L$ light curve.  Upon analysing several targets
from the large {\it HST} programme, the 1-second strategy to avoid the faint exposure systematic appears to
have mixed results.  For both WASP-19b, WASP-12b, and here for WASP-31b the second exposures
of each spacecraft orbit do not generally show the first-exposure trend as
intended, though the exposures of HAT-P-1b were still affected.
Though the 1-second strategy may not always be effective, especially
for brighter targets, it still appears to be a recommended practice for STIS programmes requiring high photometric accuracy
and long exposures as it can often save useful observing time as was usually the case here.

%%%%%%%%%%%%%%%%%%%%%%%%%%%%%%%%%%%%%%%%%%%%%%%%%%%%%%%%%%%%

\subsection{\emph{Hubble Space Telescope} WFC3 spectroscopy}

In addition to the STIS data, observations of WASP-31b were also
conducted in the infrared with Wide Field Camera 3 (WFC3) on the {\it
  HST}.  Observations began on 2012 May 13
at 12:53 using the IR $G$141 grism in forward spatial scan mode over 5 {\it HST}
orbits.  The spacial scanning is done by slewing the telescope in the cross-dispersion
direction during integration in a similar manner for each exposure, which increases the duty cycle and
greatly increases the number of counts obtained per exposure.  We used a 256 $\times$ 256 pixel subarray
with the SPARS25 NSAMP=8 readout sequence, which gives exposure times
of 134.35 s.  Our observations were conducted with a spatial scan rate of
$\sim$0.15 pixels per second, where 1 pixel = 0.13 arcsec, with the final
spatially scanned spectrum spanning $\sim$20 pixels.   The $G$141 grism images contain only the 1st order
spectrum, which spans 143 pixels with a dispersion of 4.63 nm/pixel in
the direction of the $x$-axis.   The resultant spectra
have average count rates of about 3.8$\times10^4$ electrons per pixel. 

We used the ``{\it ima}" outputs from the CALWFC3 pipeline.
For each exposure, \emph{Calwf3} conducts the following processes:
reference pixel subtraction, zero-read and dark current subtraction,
and a non-linearity correction; the resultant images are in units of
electrons per second. 
For the spectral extraction, we trimmed a 152$\times$80 box around
each spectral image and the spectra were extracted using custom
{\small IDL}
routines, similar to {\small IRAF}'s {\small APALL} procedure, with an aperture of
$\pm$12 pixels determined by minimising the standard deviation of the fitted white-light curve. The 
aperture is traced around a computed centring
profile, which was found to be consistent in the $y$-axis with an error
of 0.06 pixels. Background subtraction was applied using a clean region of
the untrimmed image, with background accounting for 0.4\% of the
  total flux measured in each wavelength column.  Subsequent data
analysis was conducted with the first orbit removed (16 exposures), as it suffers from significant thermal breathing systematic effects
similar to the STIS data. 
	
For wavelength calibration, direct images were taken in the $F$139$M$
narrow band filter at the beginning of the observations.  We assumed
that all pixels in the same column have the same effective wavelength,
as the spatial scan varied in the $x$-axis direction by less than one pixel,
resulting in a spectral range from 1.1 - 1.7~$\mu$m.  
This wavelength range is later restricted for the spectroscopic light
curve fits as the strongly sloped edges of the grism response result
in much lower S/N light curves.

%%%%%%%%%%%%%%%%%%%%%%%%%%%%%%%%%%%%%%%%%%%%%%%%%%%%%%%%%%%%

\subsection{\emph{Spitzer} IRAC photometry}

\begin{figure*}
{\centering
  \includegraphics[width=0.73\textwidth,angle=0]{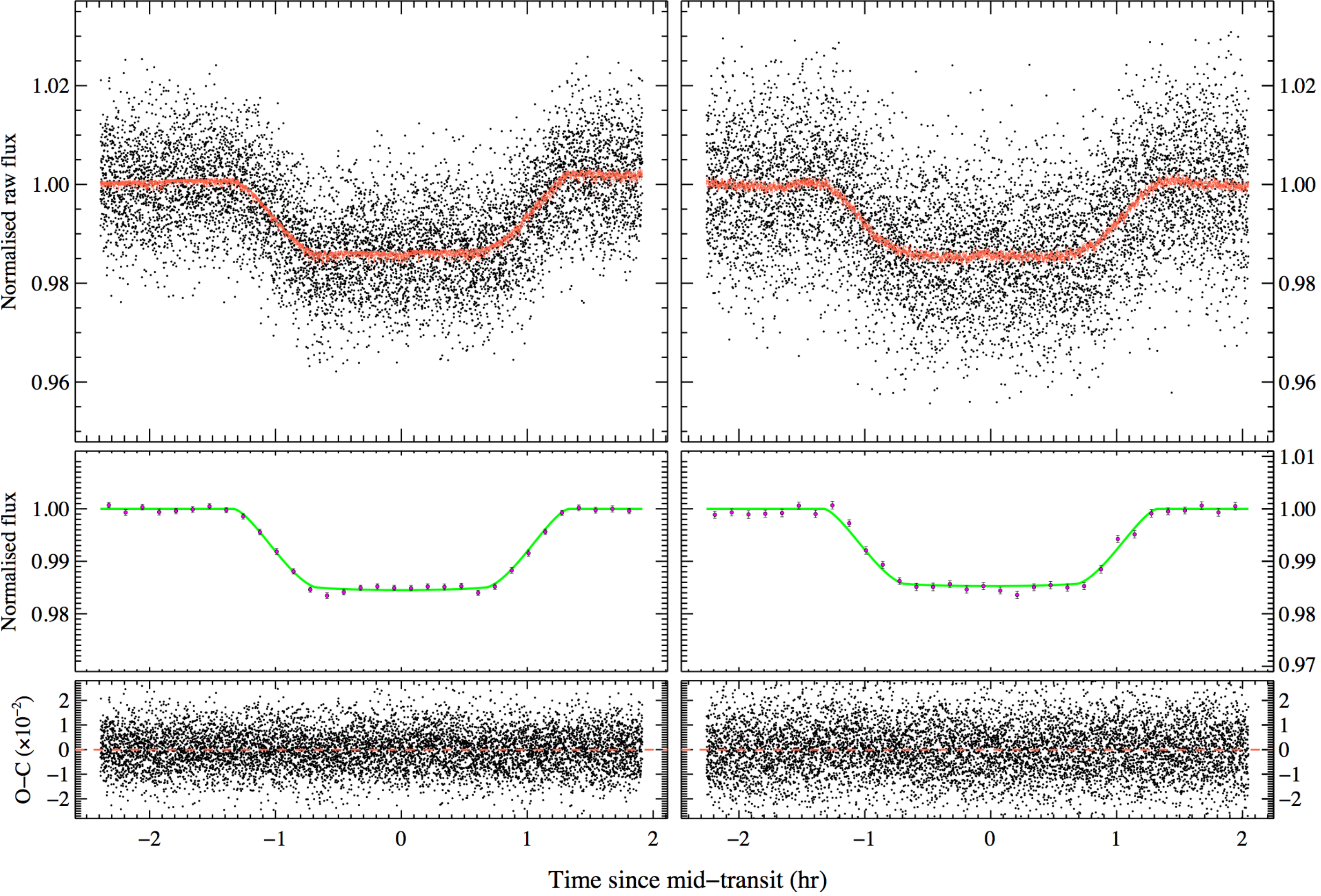}}
\caption[]{Spitzer photometry of WASP-31b for the (left) 3.6~$\mu m$
  and (right) 4.5~$\mu m$ channels.  Plotted is the raw flux along with
  the best fitting transit model including systematic effects (top), the flux corrected for 
  detector trends and binned by 8 min (middle), and residuals (bottom).    }
\label{Figure:Spitzer}
\end{figure*}

We analyse two transit observations obtained using the IRAC instrument
(Programme 90092 with P.I. Desert) on the Spitzer space telescope in
the 3.6~$\mu m$ and 4.5~$\mu m$ channels in subarray mode (32$\times$32
pixel, or 39$”\times39”$ centred on the planet’s host). The 3.6~$\mu
m$ observation was performed on UT 2013 March 9 (between 06:59 and
11:37) and the 4.5 observation was
performed on UT 2013 March 19 (between 12:19 and 16:58), each transit
containing 8320 images (see Fig. \ref{Figure:Spitzer}). 

Photometry was extracted from the basic calibrated data (BCD)
files. These images are stacked in FITS data cubes, containing 64
exposures taken in a sequence which were produced by the IRAC pipeline
(version S19.1.0) after dark subtraction, flat-fielding, linearization
and flux calibration. Both observations have effective integration
times of 1.92s per image.

We converted the images to photon counts
(i.e. electrons) by multiplying the images by the gain and individual
exposure times (FITS HEADER key words SAMPTIME and GAIN) and dividing by
the flux conversion factor (FLUXCONV).  Timing of each image was
computed using the UTC-based Barycentric Julian Date (\bjdutc) from
the FITS header keyword \bmjdobs, transforming these time stamps into
Barycentric Julian Date based on the BJD terrestrial time (TT) standard
\citep{2010PASP..122..935E}.  %Eastman
Both channels show a strong ramp feature at the beginning of the
data, and we elected to trim the first 22 min of data to allow the detector to stabilise.

We performed outlier filtering for hot (energetic) or cold (low count values) pixels in the data by examining the time series of each pixel. This task was performed in two passes, first flagging all pixels
with intensity 8-$\sigma$ or more away from the median value computed
for each frame from the 10 surrounding images. The values of these
flagged pixels were replaced with the local median value. In the
second pass, we flagged and replaced outliers above the 4-$\sigma$
level, following the same procedure. The total fraction of corrected
pixels was 0.26 per cent for the 3.6~$\mu$m and 0.07 per cent for the 4.5~$\mu$m.  

We estimated and subtracted the background flux from each image of the
time series. To do this we performed an iterative 3-$\sigma$ outlier
clipping for each image to remove the pixels with values associated
with the stellar PSF, background stars or hot pixels, created a
histogram from the remaining pixels, and fitted a Gaussian to determine
the sky background. 

We measured the position of the star on the detector in each image
incorporating the flux-weighted centroiding method\footnote{as
  implemented in the {\small IDL centroider.pro}, provided on the
  \sp~home-page.} using the background subtracted pixels from each
image for a circular region with radius 3 pixels centred on the
approximate position of the star.  While we could perform PSF
centroiding using alternative methods such as fitting a
two-dimensional Gaussian function to the stellar image, previous
experiences with warm \sp~photometry showed that the flux-weighted
centroiding method is either equivalent or superior 
\citep{2011ApJ...727...23B,2012ApJ...754...22K, 2013ApJ...766...95L}.
% Beerer et al. 2011, Knutson et al. 2012, Lewis et al. 2013). 
The variation of the $x$ and $y$ positions of the stellar psf on the
detector were 0.17 and 0.18 pixels for the $3.6 \mu m$ channel and
0.76 and 0.24 pixels for the $4.5 \mu m$ channel. 

We extract photometric measurements from our data following two
methods. First, aperture photometry was performed with IDL routine
aper using circular apertures ranging in radius from 1.5 to 3.5 pixels
in increments of 0.1. The best result was selected by measuring the
flux scatter of the out-of-transit portion of the light curves for
both channels after filtering the data for 5-$\sigma$ outliers with a
width of 20 data points.  

In addition we also performed photometry with a time-variable aperture,
adjusting the size of the aperature for each image with the noise pixel parameter
\citep{2005MNRAS.361..861M,2012ApJ...754...22K}, %Mighell Knutson et
which is described in Section 2.2.2 of the IRAC instrument handbook
and has been used in previous exoplanet studies to improve the results
of warm Spritzer photometry (e.g. 
\citealt{2012ApJ...754...22K,2014ApJ...781..109O}).   %, Knutson et al 2012, Orourke et al. 2013. 
The noise pixel parameter is proportional to the FWHM of the stellar point-spread function squared
\citep{2005MNRAS.361..861M} %Might 2005
and is defined as:
\begin{equation}
\tilde{\beta} = \frac{(\sum_{i} I_i)^2}{ \sum_{i} I_i^2},
\end{equation}
where $I_i$ is the intensity detected by the $i$th pixel. We use each
image to measure the noise pixel value, applying an aperture with a radius 4
pixels, ensuring that each pixel is considered if the border of
the aperture crosses that pixel. We extracted source fluxes from both
channel time series for aperture radii, $r$, following the relation: 
\begin{equation}
r = \sqrt{\tilde{\beta}} a_{0} + a_{1},
\end{equation}
where $a_{0}$ and $a_{1}$ were respectively varied in the ranges 0.8
to 1.2 and -0.4 to +0.4 with a step size of 0.1.

\begin{figure}
 {\centering
  \includegraphics[width=0.47\textwidth,angle=0]{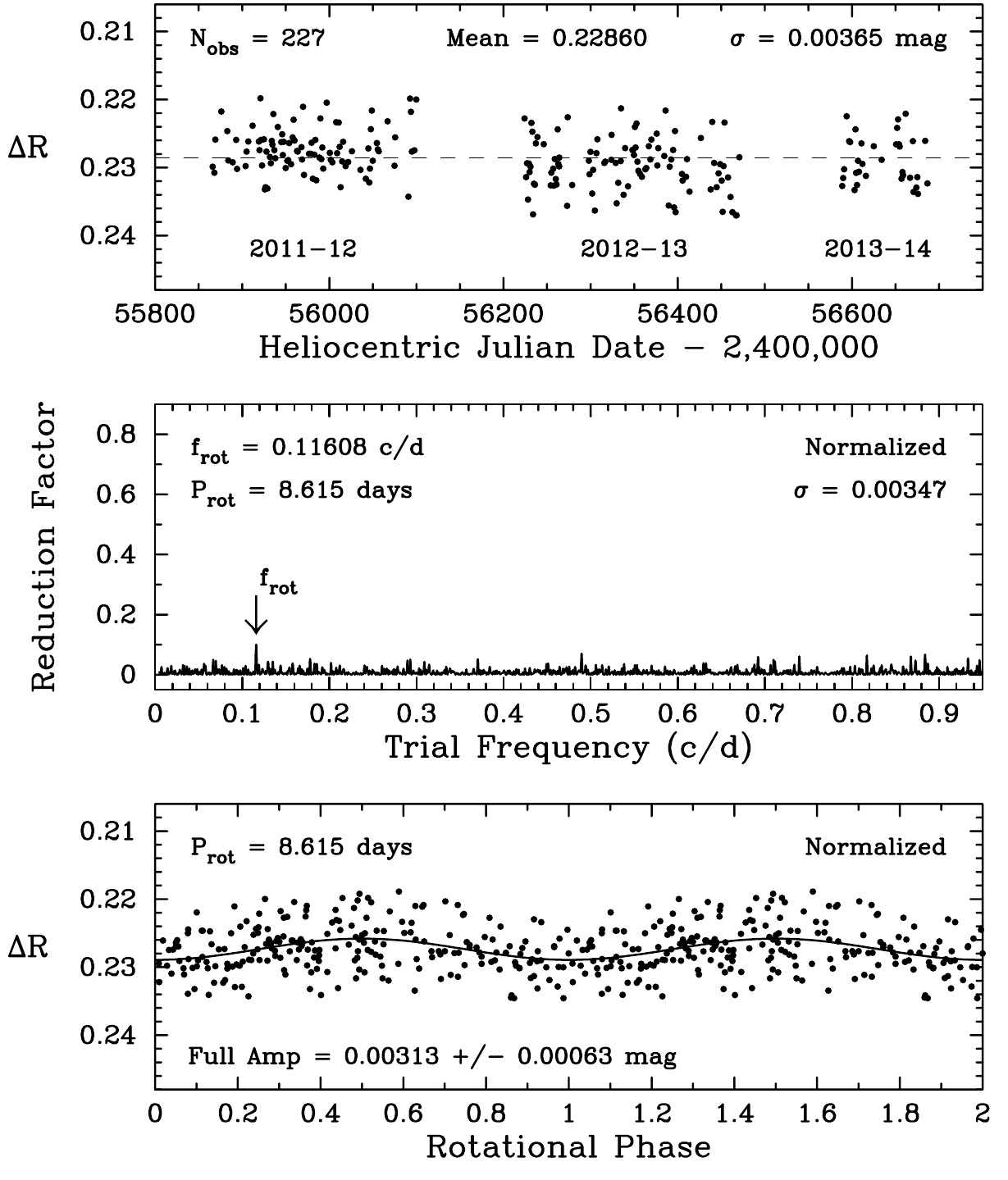}}
\caption[]{$Top$: The composite nightly Cousins $R$ band photometric data set 
of WASP-31 from the 2011--2012, 2012--2013, and 2013--2014 observing seasons, 
acquired with the C14 Automated Imaging Telescope at Fairborn Observatory.  
The data within each observing season scatter from night to night with a
standard deviation between 0.003 and 0.004 mag.  This is the approximate
limit of our measurement precision.  The scatter of the yearly means is only 
0.001 mag. $Middle$: The frequency spectrum of the composite data set, 
normalized to the first seasonal mean, suggests low-amplitude variability 
at about $8.6$ d. $Bottom$: The normalized data set phased to the 
8.6 day period, which we interpret as the stellar rotation period. A 
least-squares sine fit to the 8.6-day rotation period gives a 
peak-to-peak amplitude of 0.003~mag.
}
\label{Figure:PhotM}
\end{figure}

%%%%%%%%%%%%%%%%%%%%%%%%%%%%%%%%%%%%%%%%%%%%%%%%%%%%%%%%%%%%
\subsection{Photometric Monitoring for
Stellar Activity}
\label{section:activity}

We obtained nightly photometry of WASP-31 to monitor and characterize the stellar activity over the past three observing
seasons with the Tennessee State University Celestron 14-inch (C14) 
Automated Imaging Telescope (AIT) located at Fairborn Observatory in Arizona 
\citep[see, e.g.,][]{1999PASP..111..845H,2003ASSL..288..189E}.  % Henry, Eaton 
The AIT uses an SBIG STL-1001E CCD 
camera and exposes through a Cousins $R$ filter.  Each nightly observation 
consisted of 4--10 consecutive exposures on WASP-31 and several comparison 
stars in the same field of view.  The individual nightly frames were
co-added and reduced to differential magnitudes in the sense WASP-31
minus the mean brightness of six constant comparison stars.  Each nightly observation 
has been corrected for bias, flat-fielding, pier-side offset, and for 
differential atmospheric extinction.

A total of 227 successful nightly observations (excluding a few isolated 
transit observations) were collected between 2011 October 31 and 2014 
January 28.  The individual observations are plotted in the top panel 
of Fig.~\ref{Figure:PhotM} and summarised in Table~\ref{Table:PhotM}.  The standard deviations of single 
observations, with respect to their corresponding seasonal means, range 
between 0.0031 and 0.0038 mag for the three observing seasons (Table \ref{Table:PhotM}; 
column 4).  This is near the limit of our measurement precision with the 
C14, as determined from the constant comparison stars in the field.  The 
seasonal means given in column 5 agree to within a standard deviation of only 
0.001 mag. Since we do not standardise the ensemble differential magnitudes,
this is probably consistent with the absence of year-to-year variability.

\begin{table} %==================================================
\caption{Summary of WASP-31 AIT Photometric Observations}
\label{Table:PhotM}
\begin{centering}
\renewcommand{\footnoterule}{}  % to avoid a line before footnotes
\begin{tabular}{ccccc}
\hline\hline  %==================================================
 Observing &          &                   Date Range& Sigma & Seasonal Mean \\
Season      & $N_{obs}$ &HJD$-$2400000& (mag) & (mag)\\
\hline\hline  %==================================================
 2011--2012 &  95 & 55866--56099 & 0.00316 & $0.22730\pm0.00032$ \\
 2012--2013 &  98 & 56224--56470 & 0.00379 & $0.22974\pm0.00038$ \\
 2013--2014 &  34 & 56589--56686 & 0.00345 & $0.22893\pm0.00059$ \\
\hline%=========================================================
\end{tabular}
\end{centering}
\end{table}

We performed a periodogram analysis of the composite data set in the top
panel of Fig.~\ref{Figure:PhotM} after first normalizing the data by translating the 
second and third observing seasons to have the same mean as the first 
season.  To first order, this removes low-amplitude, long-term variability 
in WASP-31 or the comparison stars as well as any uncorrected year-to-year 
systematic errors.  A frequency spectrum of the normalized data set, 
computed as the reduction in variance of the data vs. trial frequency, 
is plotted in the middle panel of Fig.~\ref{Figure:PhotM}. Trial frequencies ranged between 
0.005 and 0.95 c/d, corresponding to a period range between 1 and 200 d. 
The result suggests low-amplitude brightness variability with a period of
$8.6\pm0.1$ d.  This is consistent with the stellar rotation period of 
$7.9~\pm~0.7$ d predicted by \citet{2011A&A...531A..60A} %Anderson
from their observed 
$v\sin{i}$ and computed stellar radius $R_{*}$.  Therefore, we take 8.6 d 
as the stellar rotation period made visible by rotational modulation in 
the visibility of weak surface features.  

The three-year normalized data set is plotted in the bottom panel of Fig.~\ref{Figure:PhotM}, 
phased to the rotation period of 8.6 d.  A least-squares sine fit 
gives a peak-to-peak amplitude of $0.0031~\pm~0.0006$ mag. Thus, while we 
find probable low-amplitude rotational modulation in stellar brightness, we 
find no evidence for significant variability above the noise at or near 
the planetary orbital period (corresponding to a frequency 
of 0.2936 c/d in Fig.~1). A least-squares sine fit of the normalized data
to the orbital period gives a peak-to-peak amplitude of only 
$0.0014~\pm~0.0006$ mag.
Since the 8.6-day stellar rotation period and the 3.4-day planetary orbital 
period are well separated, the radial velocity variations are clearly due
to planetary reflex motion with very little jitter caused by surface activity
\citep[e.g.,][]{2001A&A...379..279Q, 2004AJ....127.1644P}.  %Queloz, Paulson

In addition, we also measure a low stellar activity for WASP-31 as
determined by the Ca\,{\sc ii} H~\&~K lines, where we find an activity index of
$\rm{log}(R'_{HK})=-5.4$ using archival HARPS spectra.

From equation 5 of Sing et al. (2011), activity levels of 0.0031 mag
limits the effects of the photometric variability on the measured planet-to-star
radius contrast to 0.00018
$R_{p}/R_{*}$.
With low levels of stellar activity, {\it we conclude their effects
on measuring WASP-31b's transmission spectrum are minimal.}
A low activity star in this case is particularly advantageous for transmission
spectroscopy, as it minimizes or eliminates any potential uncertainty
in the interpretation of the transmission spectrum accross the
spectral regime and from different datasets and epochs.

%%%%%%%%%%%%%%%%%%%%%%%%%%%%%%%%%%%%%%%%%%%%%%%%%%%%%%%%%%%%

\section{Analysis}
\label{Sec:analysis}

\subsubsection{STIS white-light curve fits}
\label{Sec:STISwhite}
 Our overall analysis methods are similar to the {\it HST} analysis of
\cite{2011MNRAS.416.1443S, 2013MNRAS.434.3252H, 2013MNRAS.436.2956S} %Sing,Huitson
and \cite{2014MNRAS.437...46N}, which we also describe briefly here. %Sing 2011
The light curves were modelled with the analytical transit
models of \cite{
2002ApJ...580L.171M}.   %Mandel \& Agol (2002).
For the white-light curves, the 
central transit time, inclination, stellar density, planet-to-star radius
contrast, stellar baseline flux, and instrument systematic trends were
fit simultaneously, with flat priors assumed.  The period was initially fixed to a literature value, before
being updated, with our final fits adopting the value from section
\ref{Sec:ephemeris}.  Both $G$430$L$ transits were fit simultaneously with a common
inclination, stellar density and planet-to-star radius contrast.  The
results from the $G$430$L$ and $G$750$L$ white-light curve fits were then used in conjunction with literature
results to refine the orbital ephemeris and
overall planetary system properties (see section \ref{Sec:ephemeris}).  

To account for the effects of limb-darkening on the transit light
curve, we adopted the four parameter non-linear limb-darkening law, 
calculating the coefficients with {\small ATLAS} stellar models ($T_{eff}$=6250, log~$g$=4.5, [Fe/H]= -0.2) following \cite{
2010A&A...510A..21S}. %Sing 
Varying the stellar model within the known parameter range has a
negligible effect on the output limb-darkening coefficients and fit transit parameters. 

As in our past STIS studies, we applied orbit-to-orbit flux corrections by
fitting for a fourth order polynomial to the photometric time series,
phased on the {\it HST} orbital 
period.  The systematic trends were fit simultaneously with the
transit parameters in the fit.  Higher order polynomial fits were not statistically justified, based upon the
Bayesian information criteria (BIC; \citealt{Schwarz1978}).  The baseline flux level of each
visit was let free to vary in time linearly, described by two fit
parameters.   In addition, for the $G$750$L$
we found it justified by the BIC to also linearly fit for
two further systematic trends which correlated with the $x$ and $y$ detector positions of the spectra,
as determined from a linear spectral trace in IRAF.  

The errors on each datapoint were initially set to the pipeline values, which
is dominated by photon noise but also includes readout noise.
The best-fitting parameters were determined 
simultaneously with a Levenberg-Marquardt least-squares algorithm 
\citep{2009ASPC..411..251M} %Marquardt
using the unbinned data.
After the initial fits, the uncertainties for each data point were
rescaled based on the standard deviation of the residuals
and any measured systematic errors correlated in time (`red noise'), thus taking into account any underestimated
errors in the data points.  

The red noise was measured by checking whether the binned residuals followed a $N^{-1/2}$ relation, when
binning in time by $N$ points.  In the presence of red noise, the
variance can be modelled to follow a $\sigma^2=\sigma_{\rm
  w}^2/N+\sigma_{\rm r}^2$ relation,
where $\sigma_{\rm w}$ is the uncorrelated white noise component, and
$\sigma_{\rm r}$ characterizes the red noise \citep{
2006MNRAS.373..231P}.  %Pont
We find that the pipeline per-exposure errors are accurate
at small wavelength bin sizes, which are dominated by photon noise,
but are in general somewhat underestimated
at larger bin sizes.  For the $G$750$L$ white-light curve we find our
light curve precision achieves 1.3$\times$ the photon noise value with
the scatter of the residuals measured at 201 ppm, while the
$G$430$L$ achieves 1.6$\times$ the photon noise value with the residual
scatter 276 ppm.
We do not find any evidence for remaining
systematic errors, as both $G$430$L$ STIS white-light curves as well as the
$G$750$L$ curve give $\sigma_r=0$.
A few deviant points were
cut at the 3-$\sigma$ level in the residuals.

The uncertainties on the fitted parameters were calculated using the
covariance matrix from the Levenberg-Marquardt algorithm, which
assumes that the probability space around the best-fit solution is
well-described by a multivariate Gaussian distribution.  Previous
analyses of other HST transit observations 
\citep{2012ApJ...747...35B, 2013ApJ...778..183L, 2013MNRAS.436.2956S,
  2014MNRAS.437...46N, 2014arXiv1403.4602K}  %Berta, Line, Sing, Nikolov, Knutson
have found this to be a reasonable approximation.  We then
compared these uncertainties to the results of a MCMC analysis
\citep{2013PASP..125...83E}, %Eastman, Gaudi & Agol (2013) 
which does not assume any functional form for
this probability distribution.  In each case, we found equivalent
results between the MCMC and the Levenberg-Marquardt algorithm for
both the fitted parameters and their uncertainties.  Inspection of the
2D probability distributions from both methods indicate that there are
no significant correlations between the systematic trend parameters
and the planet-to-star radius contrast.

%%%%%%%%%%%%%%%%%%%%%%%%%%%%%%%%%%%%%%%%%%%%%%%%%%%%%%%%%%%%

\subsubsection{WFC3 white-light curve fits}
WFC3 is now being used extensively for transiting exoplanet work 
\citep{
2012ApJ...747...35B, %Berta
2012MNRAS.422..753G, %Gibson
2013ApJ...774...95D, %Deming
2013MNRAS.434.3252H, %Huitson
2013ApJ...779..128M, %Mandell
2013Icar..225..432S,  % Swain
2013MNRAS.435.3481W, %Wakeford
2014Natur.505...66K, %Kreidberg
2014Natur.505...69K}. %,Knutson
The data quality appears mainly affected by detector persistence, which
causes a ramp-effect at high count levels and
{\it HST} thermal-breathing variations.
Thus far, parameterized methods and various common-mode techniques
have been employed to handle the systematics, which largely appear
repeatable orbit-to-orbit.

\begin{figure}
 {\centering
  \includegraphics[width=0.47\textwidth,angle=0]{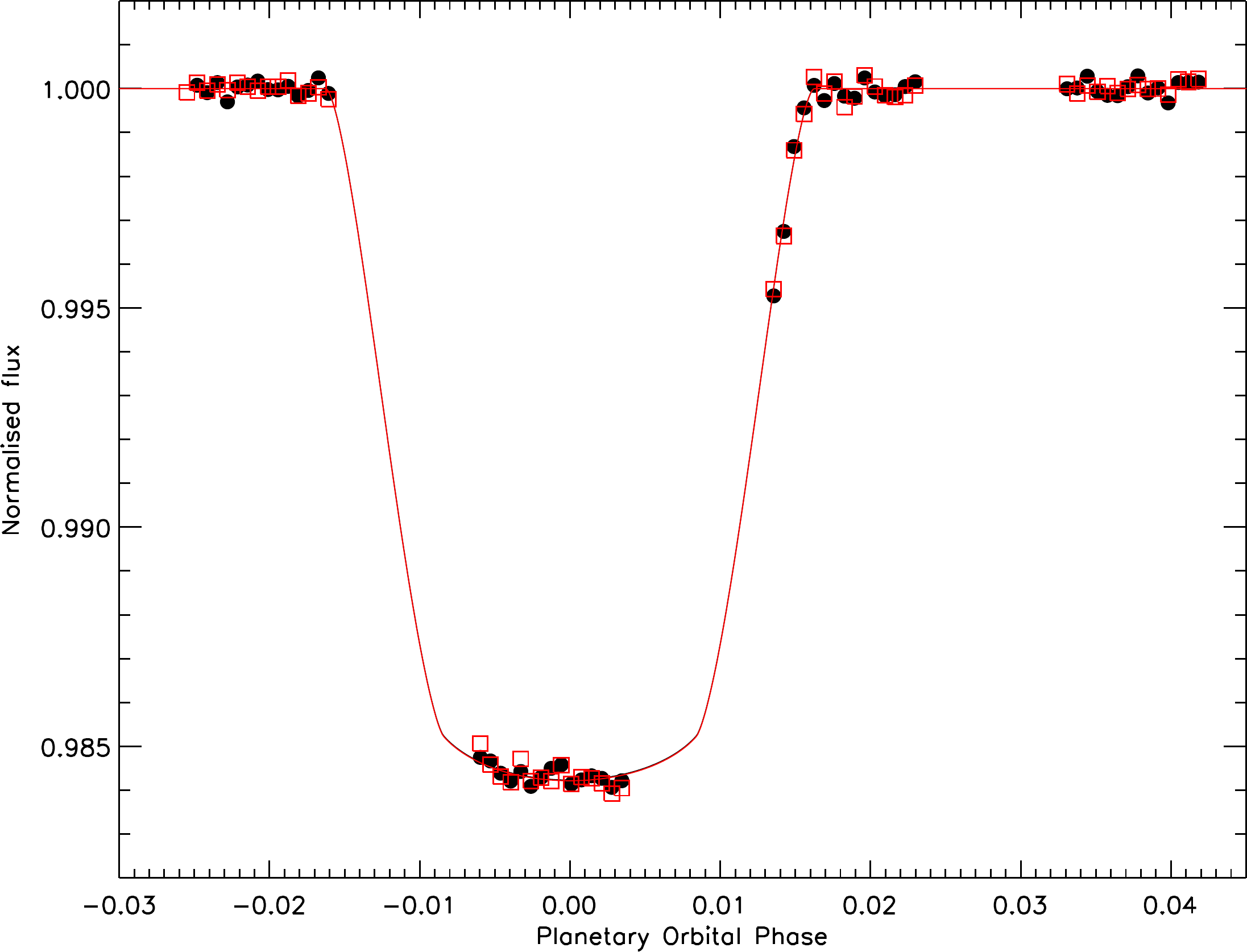}}
\caption[]{$HST$ WFC3 detrended white-light curves along with the best-fitting
  limb-darkened transit models for the cases of detrending with a polynomial of
the {\it HST} orbital phase (black, circles) and the divide-OOT
method (red, squares).  The vertical extent of the plotted symbols reflects the
flux uncertainty.}
\label{Figure:oot}
\end{figure}

Our WFC3 white-light curve transit shows small intra-orbit and orbit-to-orbit trends (see Fig. \ref{figwhite}).  We fit for the systematics errors
similarly as the STIS data, adopting a fourth order polynomial of
the {\it HST} orbital phase, a baseline flux allowed to vary linearly in time,
and a first order shift along the $y$-axis detector position as measured
through cross-correlation.  We found similar transit depths with the
divide-OOT method \citep{2012ApJ...747...35B} %Berta
but found larger overall noise properties, with the BIC favouring the
parameterized method ($\Delta$BIC=27, see Fig. \ref{Figure:oot}).  

We find our white-light curve precision achieves about 1.47$\times$ the photon noise value with
the scatter of the residuals measured at 157 ppm.  There is no evidence for remaining
systematic errors, as we estimate a value of $\sigma_r=0$ from the
binning technique.  With these precision levels, the near-IR transit
depth is measured to about a 53 ppm accuracy.  

\subsubsection{Spitzer light curve fits}
We analyse the $3.6 \mu m$ and $4.5\mu m$ channel light curves
following standard procedures for warm \sp.  First we normalize the
light curves in flux so that the out-of transit data is equal to
unity.  We employ the analytical transit models of \cite{2002ApJ...580L.171M} %Mandel
adopting the non-linear limb darkening law 
computed for the spectroscopic stellar parameters of the target (see Table 2). To
correct the \sp~systematic effects, we used a parametric model of the
form: 
\begin{equation}
f(t) = a_{0} + a_{1}x  + a_{2}x^2  + a_{3}y  + a_{4}y^2 + a_{5}xy + a_{6}t  
\end{equation}
where $f(t)$ is the stellar flux as a function of time, $x$ and $y$
are the positions of the stellar centroid on the detector, $t$ is time
and $a_{0}$ to $a_{6}$ are the free parameters in the fit 
  \citep{2009ApJ...699..478D, 2013ApJ...770..102T}.  %Desert, Todorov
Inclusion of the quadratic and cross position terms were
  justified by the BIC ($\Delta$BIC=69).
Our results are not sensitive to the inclusion of the cross-term $a_{5}xy$, though
it did help to reduce the correlated noise in channel 2.

The fits using both the time-variable and constant aperture methods
examined were examined, including 
the root-mean-square residual after fitting the light curves from each
channel as well as the white and red noise components measured using 
the wavelet method \citep{2009ApJ...704...51C}.  %Carter \& Winn 2009 
We found the 3.6~$\mu$m channel favoured the time-variable aperture
while the 4.5~$\mu m$ channel favoured a constant aperture (see
Fig. \ref{Figure:Spitzer}), with both methods giving consistent transit depths for each
channel.  To compare with the {\it HST} transmission spectra, we set the
transit ephemeris and system parameters to the values found in
Sec. \ref{Sec:ephemeris} and re-fit the light curves.  We find
residual red noise in both channels, which is comparable to the binned
white noise on the transit-duration time scale.  We estimate values for $\sigma_r$ of 
212 and 346 ppm for channels 1 and 2 respectively using either the
binning or wavelet techniques, which we account for in the final
quoted error bars.

%%%%%%%%%%%%%%%%%%%%%%%%%%%%%%%%%%%%%%%%%%%%%%%%%%%%%%%%%%%%

\subsection{System parameters and transit ephemeris}
\label{Sec:ephemeris}

\begin{figure}
 {\centering
  \includegraphics[width=0.47\textwidth,angle=0]{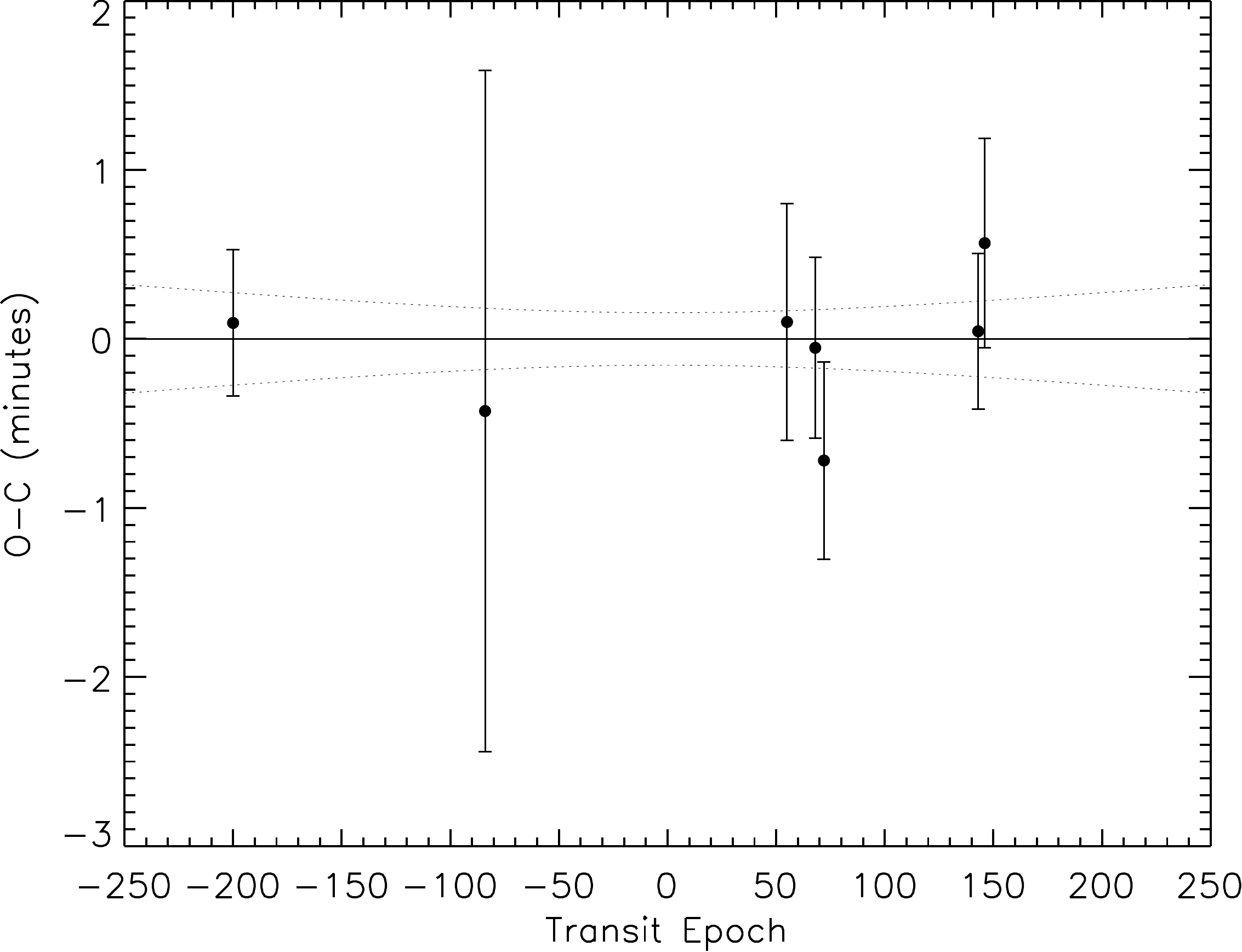}}
\caption[]{O-C diagram for the transit times of WASP-31 using the
  transits listed in Table \ref{Table:ttimes} and a linear ephemeris.
The 1-$\sigma$ error envelope on the ephemeris is plotted as the dotted lines.}
\label{Figure:Timing}
\end{figure}

We used the central transit times of the {\it HST} and {\it Spitzer} data along with the
transit times of \cite{2011A&A...531A..60A}
\cite{2011AJ....142..115D}
to determine an updated transit ephemeris.  
The transit times are listed in Table \ref{Table:ttimes}, shown
in Fig. \ref{Figure:Timing} and fit with a linear function of the
period $P$ and transit epoch $E$,
\begin{equation} T(E) = T(0) + EP. \end{equation}
We find a
period of $P = 3.40588600\pm0.00000078$ (d)
and a mid-transit time of $T(0)=2455873.86662\pm$0.00011 (BJD).
A fit with a linear ephemeris is a good fit to the data,
having a $\chi^2$ of 2.5 for 5 degrees of freedom.

\begin{table} %==================================================
\caption{Transit Timing for WASP-31b}
\label{Table:ttimes}
\begin{centering}
\renewcommand{\footnoterule}{}  % to avoid a line before footnotes
\begin{tabular}{rlll}
\hline\hline  %==================================================
 Epoc & BJD$_{TBD}$      &                    BJD error & Notes\\
\hline\hline  %==================================================
$-$200  &  2455192.6895      & 0.0003          & Anderson et al. (2011)\\  %\cite{}\\ %;Anderson  converted to BJD
$-$84    &  2455587.7719      & 0.0014          & Dragomir et al. (2011)\\   %\cite{};Dragomir
55          &  2456061.19042    & 0.00049        &  {\it HST} WFC3\\ 
68          &  2456105.46683    & 0.00037        & {\it HST} $G$430$L$\\ 
72          &  2456119.08991    & 0.00041        & {\it HST} $G$750$L$\\ %;HST $G$750$L$ 
143        &  2456360.90835    & 0.00032        & Spitzer 3.5~$\mu$m \\
146        &  2456371.12637    & 0.00043        & Spitzer 4.6~$\mu$m\\
\hline%=========================================================
\end{tabular}
\end{centering}

\end{table}

We also refined the transit parameters for the inclination, $i$, and stellar
density, $\rho_{*}$, using the white-light curve fits of the $G$750$L$ and $G$430$L$, in conjunction
with the two Spitzer light curves and the values of
\cite{2011A&A...531A..60A}.  %Anderson
Given that all our measurements were highly consistent (within the
errors) with both each
other and with those of \cite{2011A&A...531A..60A}, we computed the weighted mean value of
both the inclination, semi-major axis to stellar radius ratio, and the
stellar density finding values of
$i=84.67\pm0.11^\circ$, $a/R_{*}=8.19\pm0.10$, and
$\rho_{*}=0.895\pm0.033$ (cgs) respectively.

%%%%%%%%%%%%%%%%%%%%%%%%%%%%%%%%%%%%%%%%%%%%%%%%%%%%%%%%%%%%

%%%%%%%%%%%%%%%%%%%%%%%%%%%%%%%%%%%%%%%%%%%%%%%%%%%%%%%%%%%%

\begin{figure*}
 {\centering
  \includegraphics[width=0.84\textwidth,angle=0]{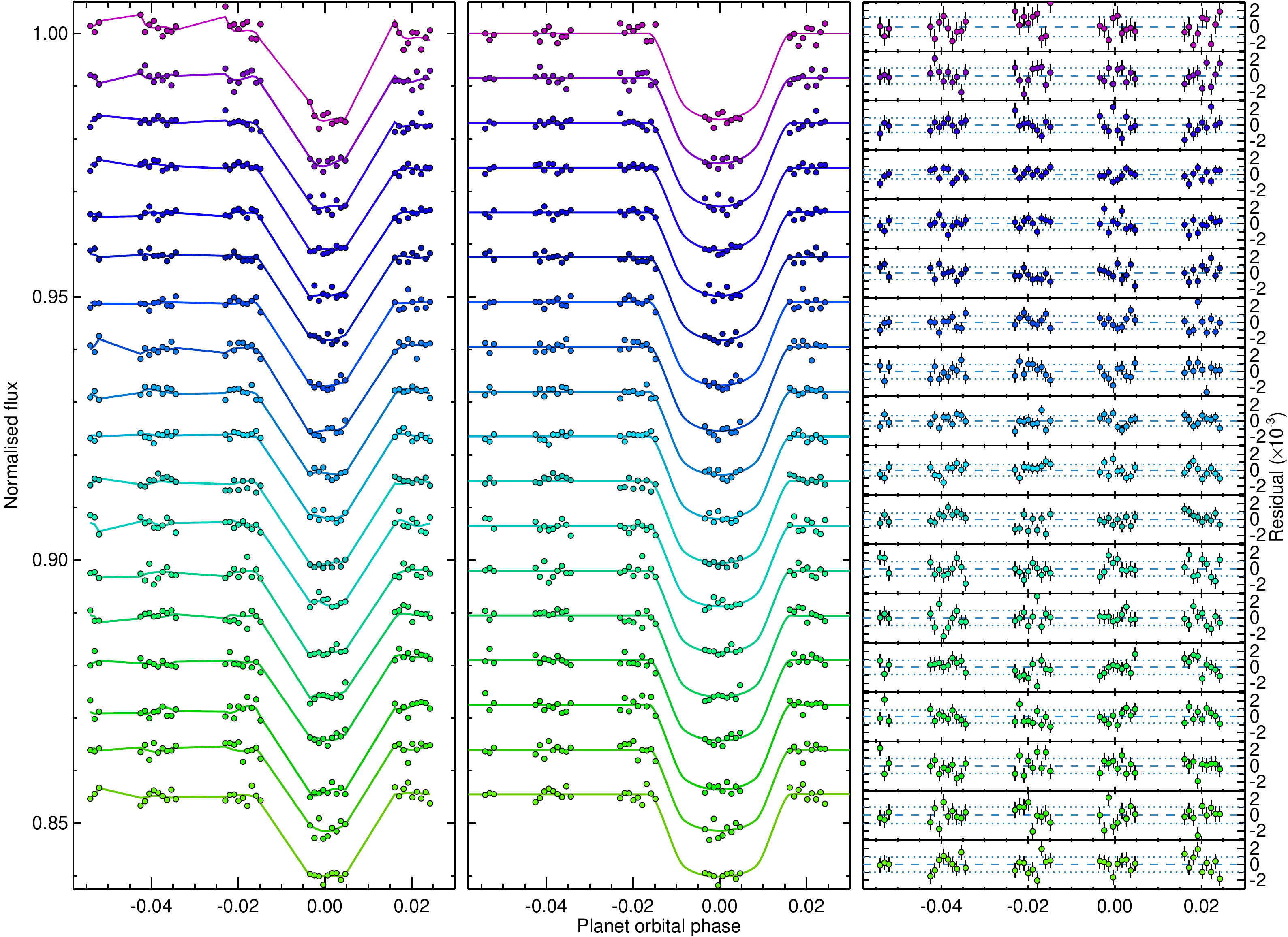}}
\caption[]{{\it HST} STIS $G$430$L$ spectral bin transit light curves for visit 15.  (Left) Light curves for the {\it HST} STIS $G$430$L$ spectral
  bins with the common-mode trends removed, over-plotted with the best fitting
  systematics model.  The points have been
  offset in relative flux, and systematics model plotted with connecting lines for clarity purposes.  The light curves are
  ordered in wavelength, with the shortest wavelength
  bin shown at the top and longest wavelength bin at the
  bottom. (Middle) Light curves fully corrected for systematic errors,
  with the best-fitting joint transit model over-plotted.  (Right) Residuals
  plotted with 1-$\sigma$ error bars along with the standard deviation
  (dotted lines).}
\label{Figure:spec$G$430$L$}
\end{figure*}
 \begin{figure*}
 {\centering
  \includegraphics[width=0.84\textwidth,angle=0]{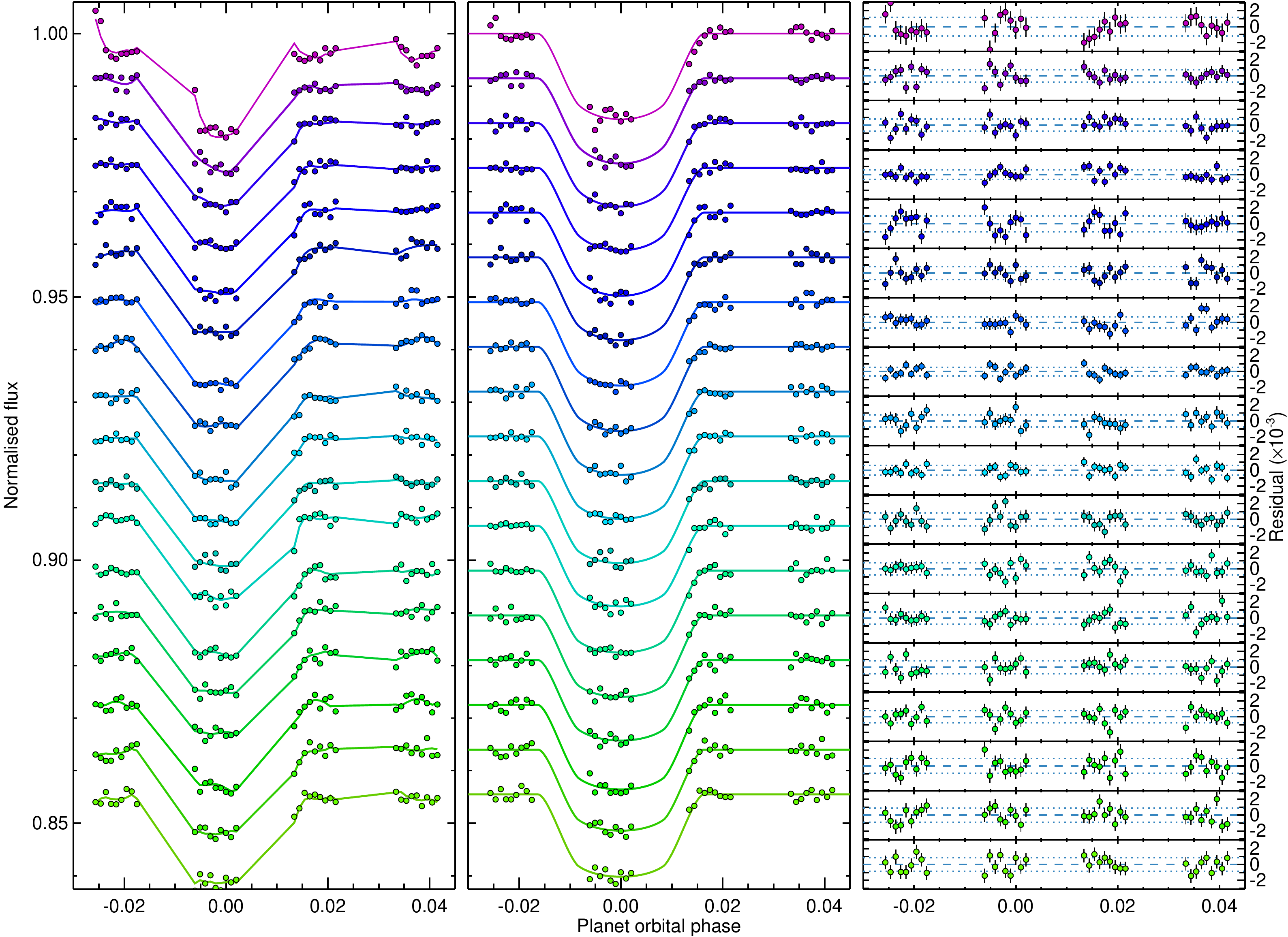}}
\caption[]{The same as Fig. \ref{Figure:spec$G$430$L$}, but for the $G$430$L$ spectral bins of
  visit 16.}
\label{Figure:spec$G$430$L$v16}
\end{figure*}
 \begin{figure*}
 {\centering
  \includegraphics[width=0.84\textwidth,angle=0]{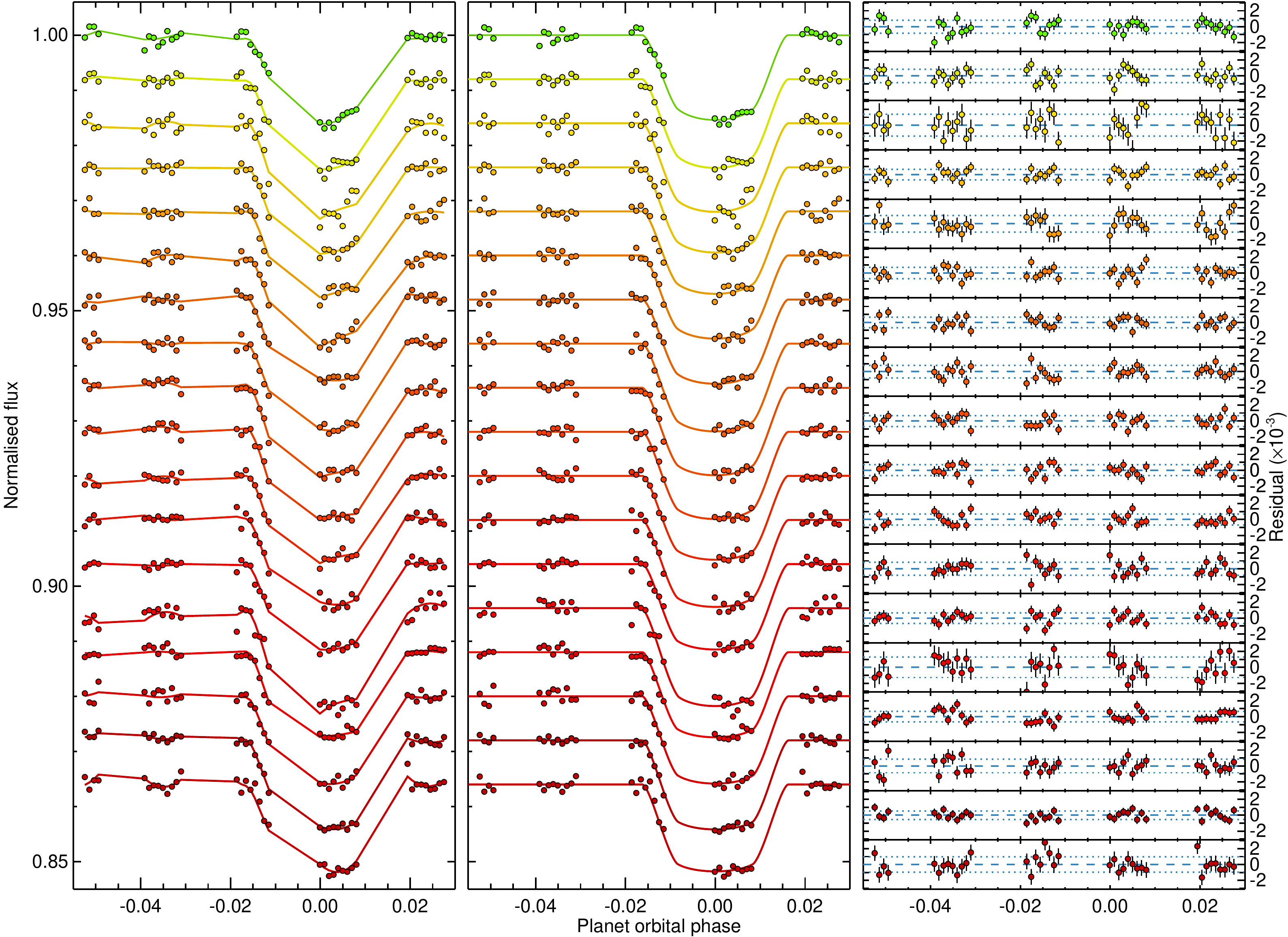}}
\caption[]{The same as Fig. \ref{Figure:spec$G$430$L$}, but for the $G$750$L$ spectral bins.}
\label{Figure:spec$G$750$L$}
\end{figure*}

%%%%%%%%%%%%%%%%%%%%%%%%%%%%%%%%%%%%%%%%%%%%%%%%%%%%%%%%%%%%
\subsection{Transmission spectrum fits}
\begin{figure*}
 {\centering
  \includegraphics[width=0.9\textwidth,angle=0]{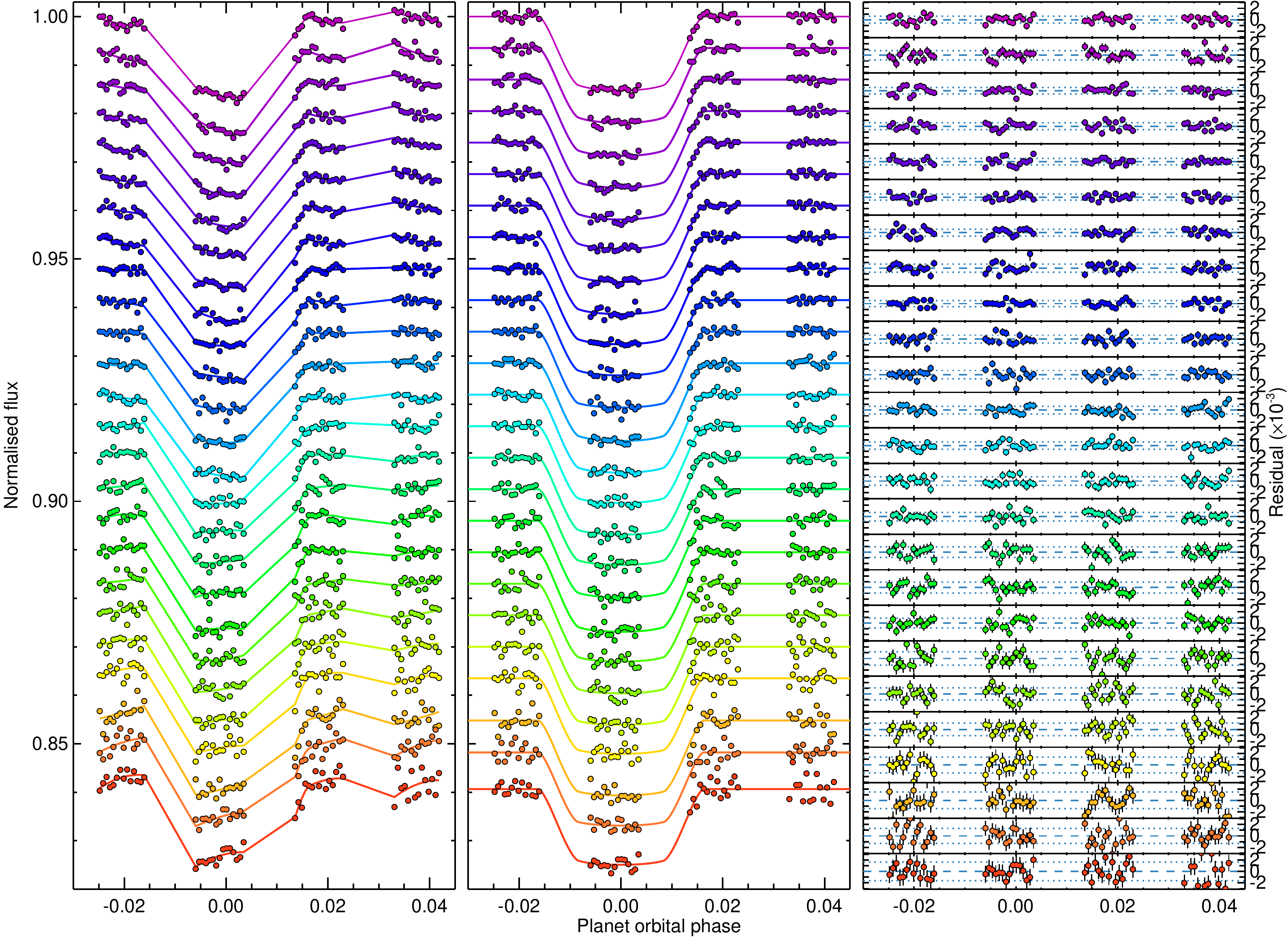}}
\caption[]{The same as Fig. \ref{Figure:spec$G$430$L$}, but for the WFC3 $G$141
    spectral bins.}
\label{Figure:specWFC3}
\end{figure*}

We extracted various wavelength bins for the STIS $G$430$L$, $G$750$L$, and
WFC3 $G$141 spectra, to create a broad-band transmission spectrum and
search for the expected narrowband Na and K features (see
Figs. \ref{Figure:spec$G$430$L$}, \ref{Figure:spec$G$430$L$v16}, \ref{Figure:spec$G$750$L$}, and
\ref{Figure:specWFC3}).  In these transit fits, we fixed the transit
ephemeris, inclination, and stellar density to the results from
Sec. \ref{Sec:ephemeris}.  The parameter $R_{P}/R_{*}$ was allowed to
fit freely, as well as the
baseline flux and model parameters describing the instrument systematic trends.  
The limb-darkening parameters were 
fixed to the model values, with the four non-linear coefficients for each bin 
individually calculated from the stellar model, taking into account
the instrument response (see Table \ref{Table:LDTrans}).
Given our overall lack of ingress or egress data, we were not able to
perform a sensitive test of the stellar limb-darkening as was done on
WASP-12A \citep{2013MNRAS.436.2956S}.  %Sing

As done for WASP-12b \citep{2013MNRAS.436.2956S}, %Sing 
we handled the modelling of systematic errors in the spectral bins by
two methods.  We allowed each light curve to be fit independently
with a parameterized model and also separately fit the spectral bins
by removing 
common-mode trends from each spectral bin, before fitting for 
residual trends with a parameterized model containing fewer free parameters.
For both the STIS and WFC3, removing the common trends significantly reduces the amplitude of the
observed systematics, as these trends are
observed to be similar wavelength-to-wavelength across a detector. 
The two methods give very similar results, though we choose the
common-mode method to quote our subsequent results as it produces slightly higher overall S/N with
more degrees of freedom (d.o.f.).
The $\Delta$BIC values are on average lower for the common-mode
method by 10.5 for the $G$430$L$ light curves, 13.1 for $G$750$L$
light curves, and 25.7 for the WFC3 light curves.

\begin{figure}
 {\centering
  \includegraphics[width=0.47\textwidth,angle=0]{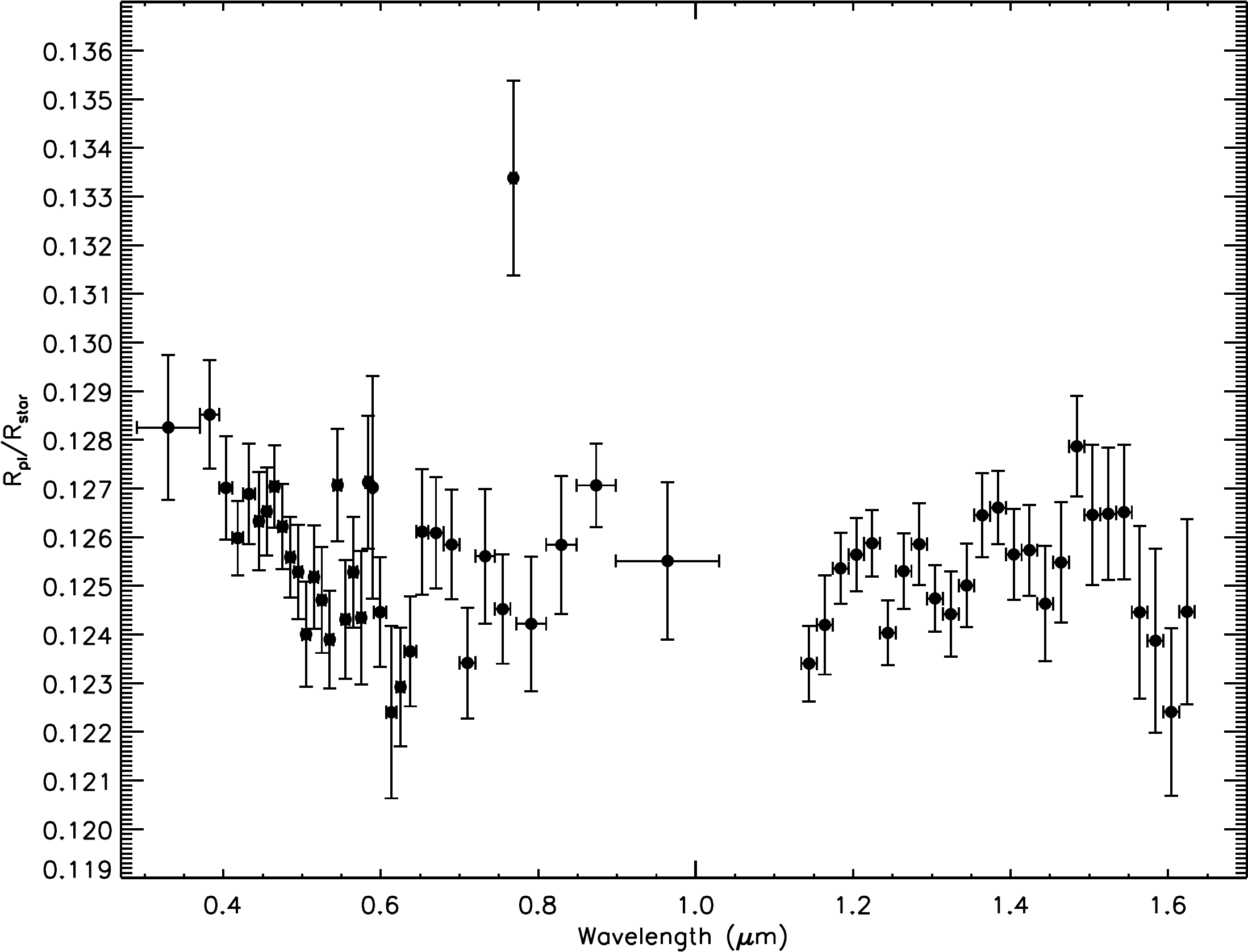}}
\caption[]{Combined {\it HST} STIS, {\it HST} WFC3 transmission
  spectrum of WASP-31b.}
\label{Figure:Broadbandspec}
\end{figure}

With the common-mode method, we reach average precisions of 88 per
cent the photon noise limit with the $G$430$L$, 92 per cent for the
$G$750$L$, and 72 per cent with the WFC3 with no significant red-noise detected in any of our spectral bins.
The transmission spectral results for both {\it HST} and Spitzer are given in Table
\ref{Table:LDTrans} and shown in Fig. \ref{Figure:Broadbandspec}.

%%%%%%%%%%%%%%%%%%%%%%%%%%%%%%%%%%%%%%%%%%%%%%%%%%%%%%%%%%%%

\section{Discussion}

WASP-31b has a zero-albedo equilibrium temperature of $T_{\rm eq}=1570$~K,
and is highly inflated with a radius of $R_{p}=1.55 R_{Jup}$, mass of
$M_{p}=0.48 M_{Jup}$, and a surface gravity of $g=4.56$ m/s$^2$ 
\citep{2011A&A...531A..60A}. %Anderson
At these temperatures, cloud-free solar-abundance models predict
significant optical absorption by Na and K atoms, covering most of the
optical spectrum with pressure-broadened line wings, with the near-infrared dominated
by H$_2$O \citep{2010ApJ...719..341B, %Burrows
2010ApJ...709.1396F}.   %Fortney

With a very low surface gravity, WASP-31b has large predicted
transmission spectral features, which can be estimated from the
analytical relation for the wavelength-dependant transit-measured altitude $z(\lambda)$ of a hydrostatic atmosphere
found in \cite{2008A&A...481L..83L}, %Lecav.
\begin{equation} 
\label{Eq:Lecav}
z(\lambda)=H \mathrm{ln} \left(
    \frac{\varepsilon_{\rm abs}  P_{\rm ref}
      \sigma(\lambda)}{\tau_{\rm eq}}
    \sqrt{\frac{2 \pi R_{\rm p}}{kT\mu g}} \right), \end{equation}
where 
$\varepsilon_{abs}$ is the abundance of dominating absorbing species, 
$T$ is the local gas temperature,
$\mu$ is the mean mass of the atmospheric particles,
$k$ is Boltzmann's constant,
$P_{\rm ref}$ is the pressure at the reference altitude,
$\sigma(\lambda)$ is the absorption cross-section, and
$H=kT/\mu g$ is the atmospheric scale height.
For WASP-31b, the scale height is estimated to be 1220 km at 1550 K
(or 0.0014 $R_{P}/R_{*}$), which gives expected Na, K, and H$_2$O transmission spectral features which are $\sim$0.006
$R_{P}/R_{*}$ in amplitude.  Our transmission spectrum compares very
favourably to these predicted signals, as our average per-bin
precisions are 0.0011$R_{P}/R_{*}$, which is less then one scale height.

%%%%%%%%%%%%%%%%%%%%%%%%%%%%%%%%%%%%%%%%%%%%%%%%%%%%%%%%%%%%
\subsection{Interpreting the transmission spectrum}

The broad-band transmission spectrum of WASP-31b seen in
Fig. \ref{Figure:Broadbandspec} has three distinct
features:  a significant blue-ward slope shortward of $\sim$0.6~$\mu$m, a largely
flat spectrum longward of about 0.6~$\mu$m which extends through the red-optical out to infrared wavelengths, and a
strong but narrow K line core.  We compare these spectral features 
to both analytic models and those from the modelling suites of 
\cite{2010ApJ...719..341B} %Burrows
and \cite{2010ApJ...709.1396F}.  %Fortney

The models from \cite{2008ApJ...678.1419F,2010ApJ...709.1396F} %Fortney
include isothermal models as well as those with a self-consistent treatment of radiative transfer and chemical equilibrium of neutral
and ionic species.  Chemical mixing ratios and opacities assume solar
metallicity and local thermo-chemical equilibrium accounting for condensation
and thermal ionisation but no photoionisation 
\citep{1999ApJ...519..793L,2002Icar..155..393L, %Lodders 1999, Lodders\& Fegley 2002;; 
2008ApJS..174..504F}.  %Freedman, Marley \& Lodders 2008

The models from \cite{2010ApJ...719..341B} %Burrows et al. (2010) 
and \cite{2012ApJ...756..176H} %Howe \& Burrows (2012) 
use a 1D day-side T–P profile with stellar irradiation, in radiative and
thermo-chemical equilibrium.  Chemical mixing ratios and opacities assume
solar metallicity and local chemical equilibrium accounting for
condensation and thermal ionisation but no photoionisation, using the
opacity data base from \cite{2007ApJS..168..140S} %Sharp \& Burrows (2007) 
and the equilibrium chemical abundances from
\cite{1999ApJ...512..843B} and \cite{2001RvMP...73..719B}.  %Burrows
%%%%%%%%%%%%%%%%%%%%%%%%%%%%%%%%%%%%%%%%%%%%%%%%%%%%%%%%%%%%
\subsubsection{A cloud deck}
\label{Sec:cloud}
\begin{figure}
 {\centering
  \includegraphics[width=0.47\textwidth,angle=0]{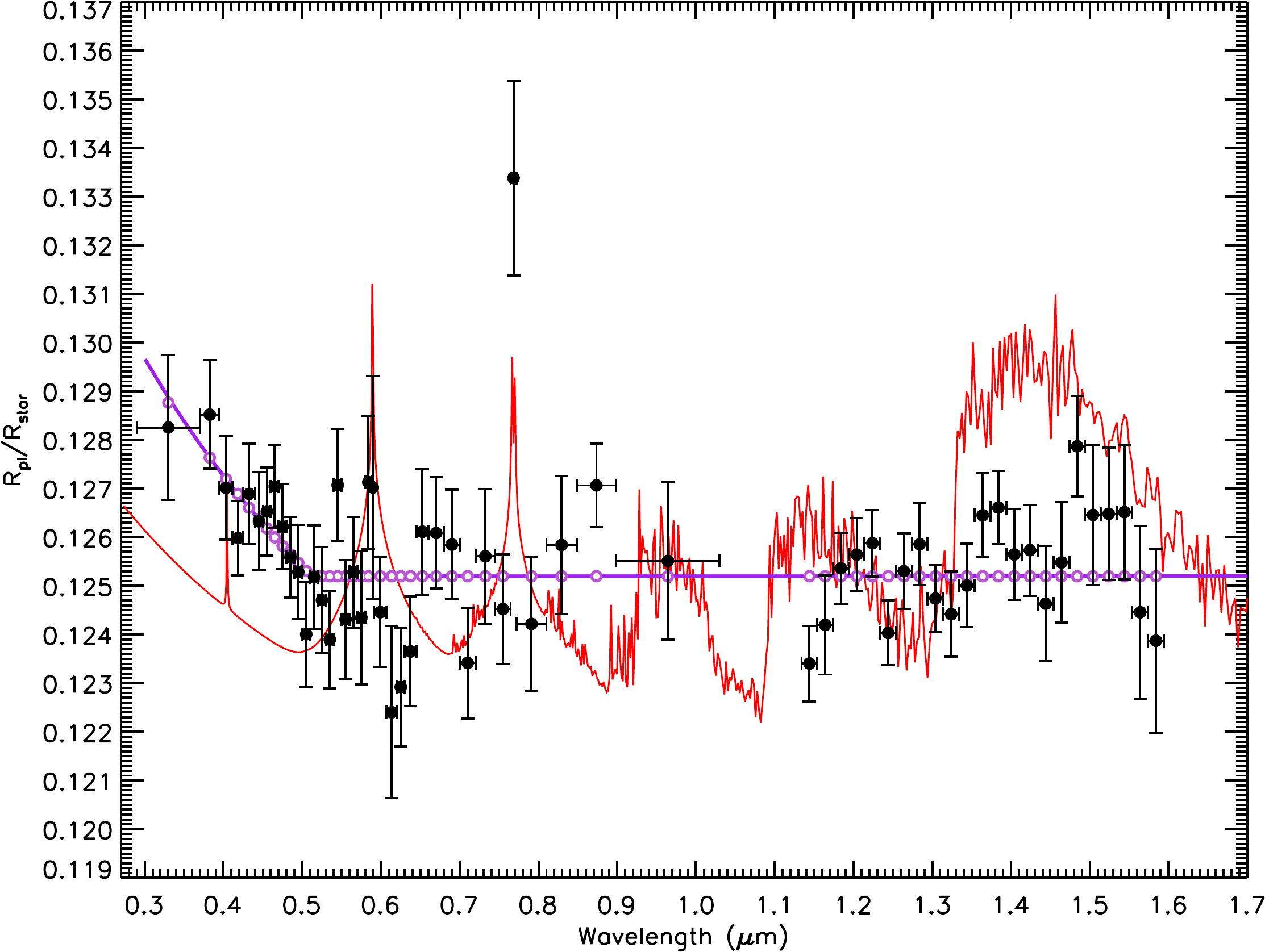}}
\caption[]{Plotted is the broadband transmission spectral data
  along with atmospheric models including a
  1500 K cloud-free solar-abundance model (red) and a Rayleigh
  scattering and cloud-deck model (purple).}
\label{Figure:K}
\end{figure}

The large pressure-broadened alkali features predicted by models with
with a clear atmosphere are not seen in the data (see Fig.
\ref{Figure:K}) and cloud-free solar-abundance models can
be ruled out.  A 1500~K isothermal model from
\cite{2008ApJ...678.1419F}, %Fortney
which contains large pressure-broadened alkali line wings, 
gives a poor fit with a $\chi^2$ value of 51.0 for 20 d.o.f. when fit
to the optical {\it HST} STIS data between the wavelengths of 0.52 and
0.9~$\mu$m (excluding the potassium core).  A flat line gives a 
better fit to the data, at 5-$\sigma$ confidence, with a $\chi^2$ value of 25.9 for 20 d.o.f.  

The H$_2$O bandhead features predicted by cloud-free models in
the WFC3 spectra are also not detected, though the data are clearly of
sufficient quality to have detected such a feature (see Fig.
\ref{Figure:K}).  In particular, a 1500~K
isothermal model predicts a broad $\Delta R_p/R_{*}=0.0058$ amplitude H$_2$O
feature between 1.25 and 1.425~$\mu$m, while the WFC3 spectra does not
significantly deviate from a flat spectrum beyond the precision level
of the datapoints which average 0.0011 R$_p$/R$_{*}$ per spectral bin.
Moreover, the 1500 K isothermal model dominated by molecular H$_2$O in
the near-IR gives $\chi^2$ value of 112 for 24 d.o.f when fit
to the WFC3 data,  compared to a $\chi^2$ value of 34 for 24 d.o.f. for a flat line.  
Thus, a water dominated atmosphere can be ruled out at 8.8-$\sigma$
confidence.

There is not a significant difference in the overall measured radius
between the 4 transits of the WFC3, STIS
$G$750$L$, and Spitzer data sets.  A single flat line fitting these the 47
datapoints for wavelengths longer than $\sim$0.52~$\mu$m (excluding the Na and K lines) results in a $\chi^2$
value of 59.8 for 46 d.o.f. and R$_p$/R$_{*}$=0.12519$\pm$0.00015.

Approximations to hazy or cloudy atmospheres were generated varying the Fortney et al. and Burrows et al. isothermal models with either an
artificially added Rayleigh scattering component, simulating a
scattering haze, or a cloud deck simulated by a grey flat line at
different altitudes.  In these models, the artificial hazes and clouds
cover up some but not all of the expected atomic and molecular gaseous
species.

While a flat line is a good fit to the flat portions of the transmission
spectrum, a low-amplitude
water feature does slightly improve the $\chi^2$ quality of the fit.  
A cloudy Fortney et al. model gives a $\chi^2$ value of 31.2
for 23 d.o.f. (fitting for the altitude of the cloud deck and overall model
altitude) when compared to the WFC3 data, compared to a flat line 
 $\chi^2$ of 34.2 for 24 d.o.f.  In addition, when fit to the STIS, WFC3 and Spitzer data for wavelengths
 $>$0.52~$\mu$m (excluding the Na and K lines) the cloudy Fortney et
 al. model gives our best fit to those data with a $\chi^2$ of 56.61
 for 44 d.o.f. and a
$R_{P}/R_{*}$=0.12497$\pm$0.00014.

%%%%%%%%%%%%%%%%%%%%%%%%%%%%%%%%%%%%%%%%%%%%%%%%%%%%%%%%%%%%
\subsubsection{Rayleigh scattering and a cloud deck}
\label{Sec:RayCl}
The data shortward of $\sim$0.55~$\mu$m shows a significant blue-ward
slope, indicative of Rayleigh scattering (see Fig. \ref{Figure:Broadbandspec}).  
To further characterize and assess the significance of the slope, we
fit the near-UV transmission spectrum assuming an atmospheric opacity 
with an effective extinction (scattering+absorption) cross-section which follows a power law
of index $\alpha$, i.e. $\sigma=\sigma(\lambda/\lambda_0)^{\alpha}$
following \cite{2008A&A...481L..83L}.
With this assumption, the slope of the transmission spectrum is then proportional to the
product $\alpha T$ given by,
\begin{equation}  
  \alpha T=\frac{\mu g}{k} \frac{{\rm d}(R_{p}/R_{*})}{{\rm d ln}\lambda}. 
\end{equation}
However, compared to previous studies of HD~189733b and WASP-12b
where this analytic method was a useful diagnostic of the observed
transmission spectral slope
\citep{2008A&A...481L..83L, 2013MNRAS.436.2956S}, %Lecav&SIng
it is not
immediately apparent in WASP-31b where the blue-ward near-UV slope
transitions to a flat spectrum.  This ambiguity can have an affect on
the inferred slope (and derived temperatures) as including more of the
flat portions of the spectrum will decrease the measured slope,
lowering the inferred temperatures.     

We elected to fit a two-component model which included both
a power law cross-section at short wavelengths, and a flat spectrum at
longer wavelengths, with the transition wavelength, $\lambda_{\rm T}$, between the two
freely fit (see Fig. \ref{Figure:K}).  Excluding only the Na and K line cores from the full
transmission spectrum, this model gave a good
fit to the 59 datapoints ($\chi^2$ of 65.4 for 56 d.o.f.), 
with $\lambda_{\rm T}=0.510\pm0.031$~$\mu$m and $\alpha T =
-9284\pm$3240~K.  The uncertainty in $\lambda_{\rm T}$ increases the error in
measuring $\alpha T$ by about 40 per cent.  The result is compatible with Rayleigh scattering,
$\alpha=-4$, which results in estimated temperatures of 2300$\pm$800
K.  While the derived temperature may appear too high to be associated
with most condensates, we note that given the large error it is still within the 1-$\sigma$ uncertainty of
the zero-albedo equilibrium temperature ($T_{\rm eq}=1570$~K) where
condensates from Fe and Mg are found.
This two-component model, simulating a short-wavelength Rayleigh haze
and a long-wavelength cloud deck, results in a much better $\chi^2$ fit compared to a simple flat
line over the same wavelength range ($\chi^2$ of 87.17 for 57 d.o.f.).
Thus, the presence of a near-UV Rayleigh slope is significant at about 
4-$\sigma$ confidence.

%%%%%%%%%%%%%%%%%%%%%%%%%%%%%%%%%%%%%%%%%%%%%%%%%%%%%%%%%%%%
\subsubsection{Potassium}

In addition to measuring the broad-band transmission spectrum, we also
searched the STIS data for narrow-band signatures, notably the line
cores of the Na~{\small I} and K~{\small l} D lines and the
H$_{\alpha}$ line.  No significant absorption was found
for either Na~{\small I} or H$_{\alpha}$.  We quote a 34~\AA\ wide bin across the Na
line in our transmission spectrum for comparison purposes (see Table \ref{Table:LDTrans}), choosing
this width based on the HAT-P-1b Na detection \citep{2014MNRAS.437...46N}. %Nikolov

While no significant Na~{\small I} absorption was detected, there is a strong K line evident
in the data (see Fig. \ref{Figure:K}).  Searching both on and around the wavelength range of the K~{\small I}
doublet (with line centres at 7664 and 7699 \AA), we measured the transit depth in various bin sizes ranging from one
resolution element ($\sim$10 \AA) up to several hundred \AA, 
comparing the transit depths to the surrounding continuum regions
and that of the white-light curve.  We find a radius of
$R_{p}/R_{*}$=0.1334$\pm$0.0020 in a 78 \AA\ band (7645 and 7720
\AA) which encompasses the K~{\small I} doublet gives an optimal detection, which is larger than the
white-light radius at 4.2-$\sigma$ confidence ($\Delta
R_{p}/R_{*}$=0.0086$\pm$0.0020).  A confirmation of the  K~{\small I}
detection is provided by the fact that each line of
the potassium doublet is also individually detected, with a band from
7645 to 7683 \AA\ giving radius of $R_{p}/R_{*}$=0.1342$\pm$0.0027 and a
band from 7683 to 7720 \AA\ giving a radius of
$R_{p}/R_{*}$=0.1325$\pm$0.0032.
No significant signals were detected outside the 7645 to 7720 \AA\
region confining the potassium feature to a 78~\AA\ region, as wider bins dilute the measured potassium
absorption depth.  In addition, no other similar features rising to significant
levels above the broad-band continuum are seen at any other wavelengths in the $G$750$L$ data. 

%%%%%%%%%%%%%%%%%%%%%%%%%%%%%%%%%%%%%%%%%%%%%%%%%%%%%%%%%%%%

\subsection{Limits on the pressure level of the cloud deck}
The pressure level of the broad-band spectrum can potentially be constrained by the shape of
the potassium line and the presence (or lack) of pressure-broadened wings.  
The maximum expected pressures probed would correspond to the near-UV slope being due
to gaseous H$_2$ Rayleigh scattering, which following the fits from
section \ref{Sec:RayCl} and
\cite{2008A&A...481L..83L} %Lecavelier
would put the cloud deck radius of $R_{P}/R_{*}$=0.12497 at pressures
of 40 mbar.  However, this scenario is unlikely due to the lack of
pressure-broadened K wings, and the fact that both Na and H$_2$O would
have to both have sub-solar abundances depleted by about a factor of
100 to hide both features. 

To further constrain the pressures and abundances, we model the alkali metal lines following
\cite{2000ApJ...531..438B}, %Burrows
\cite{2005A&A...436..719I},  %Iro
and
\cite{2012MNRAS.422.2477H}.  %Huitson 
Statistical theory predicts the collisionally-broadened alkali line
shapes which vary as $(\nu-\nu_0)^{-3/2}$ outside of the impact
region, which lies between the line centre frequency $\nu_{0}$
and a detuning frequency $\Delta \sigma$ away from $\nu_{0}$.  The power-law line shape is then
truncated by an exponential cutoff term of the form
$e^{-qh(\nu-\nu_0)/kT}$, 
where $h$ is Planks constant and $q$ a parameter of order unity.
For the potassium doublet, we estimate $\Delta \sigma$ 
following \cite{2000ApJ...531..438B} using %Burrows
\begin{equation}  
\Delta \sigma = 20(T/500~{\rm K})^{0.6}~{\rm cm}^{-1}.
\end{equation}
Assuming a 1570~K equilibrium temperature for WASP-31b gives a detuning
frequency of 23\AA\ from the K line centres.  The
K doublet is spaced by 34.1 \AA, giving a total impact
region encompassing the two doublets which is 80.1\AA\ wide (7641.6 to 7722\AA).  

Within $\Delta \sigma$ of the line centre in the impact region,
we use a Voigt profile, $H(a,u)$, which is defined in terms of the
Voigt damping parameter $a$ and frequency offset $u$. 
 The frequency offset is calculated as 
$u=(\nu-\nu_0)/\Delta\nu_D$, where $\Delta\nu_D$ is the Doppler width 
given by $\Delta \nu_{\rm D} = \nu_{0}/c \sqrt{2kT/\mu_{\rm K}}$,
with $c$ the speed of light and $\mu_{\rm K}$ the mean molecular
weight of potassium.
The damping parameter is given by $a=\Gamma / (4\pi\Delta\nu_{\rm D})$,
where the transition rate $\Gamma$ is calculated following
\begin{equation}  
\label{Eq:gamma}
\Gamma=\gamma+\Gamma_{\rm col},
\end{equation}
where $\gamma$ is the spontaneous decay rate and $\Gamma_{\rm col}$ is the
half-width calculated from classical impact theory.  Assuming a van der
Waals force gives  
$\Gamma_{\rm col}=0.14(T/2000)^{-0.7}$~cm$^{-1}$ atm$^{-1}$ for K \citep{2000ApJ...531..438B,2005A&A...436..719I}.  %Burrows&iro
The cross-sections for both the sodium and potassium doublets in the
impact region are then calculated as,
\begin{equation} 
\sigma(\lambda)=\frac{\pi e^2}{m_e c}\frac{f}{\Delta\nu_D \sqrt{\pi}}H(a,u),
\end{equation}
where $f$ is the absorption oscillator strength of the spectral line,
$m_e$ is the mass of the electron and $e$ the electron charge.  With
the cross-sections, the transmission spectrum is then calculated with
equation \ref{Eq:Lecav}. 

In addition to the cloud-free solar-abundance models which can be
ruled out at high confidence (see section \ref{Sec:cloud}), we further test for the presence of collisionally broadened wings by using the
Rayleigh and cloud analytic model from section \ref{Sec:RayCl}, fixing
both $T$ and $\lambda_{\rm T}$ while adding a potassium line and adjusting
$P_{\rm ref}$ via the Rayleigh slope opacity.  This also tests the
hypothesis that the near-UV slope is due to gaseous H$_2$ Rayleigh
scattering (as opposed to aerosols), but a low-altitude cloud is still present.  With a solar
abundance of potassium and assuming H$_2$ Rayleigh
scattering, the model gives a fit with a $\chi^2$ of 71.3 with 60 d.o.f.
The model reproduces the height of the K core measurement and still contains significant collisional broadening as the presumed low-altitude cloud does not fully cover the
broadened alkali wings.
Assuming instead the Rayleigh scattering is from an aerosol (residing at higher altitudes and lower pressures) and maintaining the amplitude of the
potassium line gives a $\chi^2$ of 65.2 for 60 d.o.f., with
the model lacking significant collisionally broadened wings.  In conjunction with the results of section \ref{Sec:cloud},
pressure-broadened potassium wings can be fully excluded from
cloud-free models, and further at about the 98 per cent
confidence level for the case of H$_2$ Rayleigh
scattering with a low-altitude cloud.  We note that this result can be expected from
statistical theory, as the width of the potassium line calculated assuming
pressures of zero in equation \ref{Eq:gamma} (i.e. only natural
broadening) match very well with the measured width of the potassium line.

We estimate the maximum pressure level of the cloud deck by raising
the Rayleigh scattering opacity and K abundance by the minimum amounts
needed to maintain the K line amplitude and produce a sufficiently narrow width.   
A cloud deck at pressures below about 10 mbar reduce the
broadeneing such that the model gives a $\Delta\chi^2$ of 1 compared the
best fit model, which corresponds to a minimum aerosol opacity which is
about 4$\times$ that of gaseous H$_2$.  

As the 1.4~$\mu$m H$_2$O feature is strong, the pressure level
of the cloud deck estimated by the lack of alkali broadened wings (10
mbar) would still not be sufficient to cover H$_2$O at the mixing
ratios predicted when assuming chemical equilibrium and solar-abundances.  Based on the Fortney et al. and Burrows et al. models, we estimate 
that a cloud deck at pressures of about 1 mbar would be needed to
cover all but the very peak of the 1.4~$\mu$m H$_2$O feature at mixing
ratios assuming solar-abundances.

%%%%%%%%%%%%%%%%%%%%%%%%%%%%%%%%%%%%%%%%%%%%%%%%%%%%%%%%%%%%

%%%%%%%%%%%%%%%%%%%%%%%%%%%%%%%%%%%%%%%%%%%%%%%%%%%%%%%%%%%%

\begin{figure*}
 {\centering
  \includegraphics[width=0.94\textwidth,angle=0]{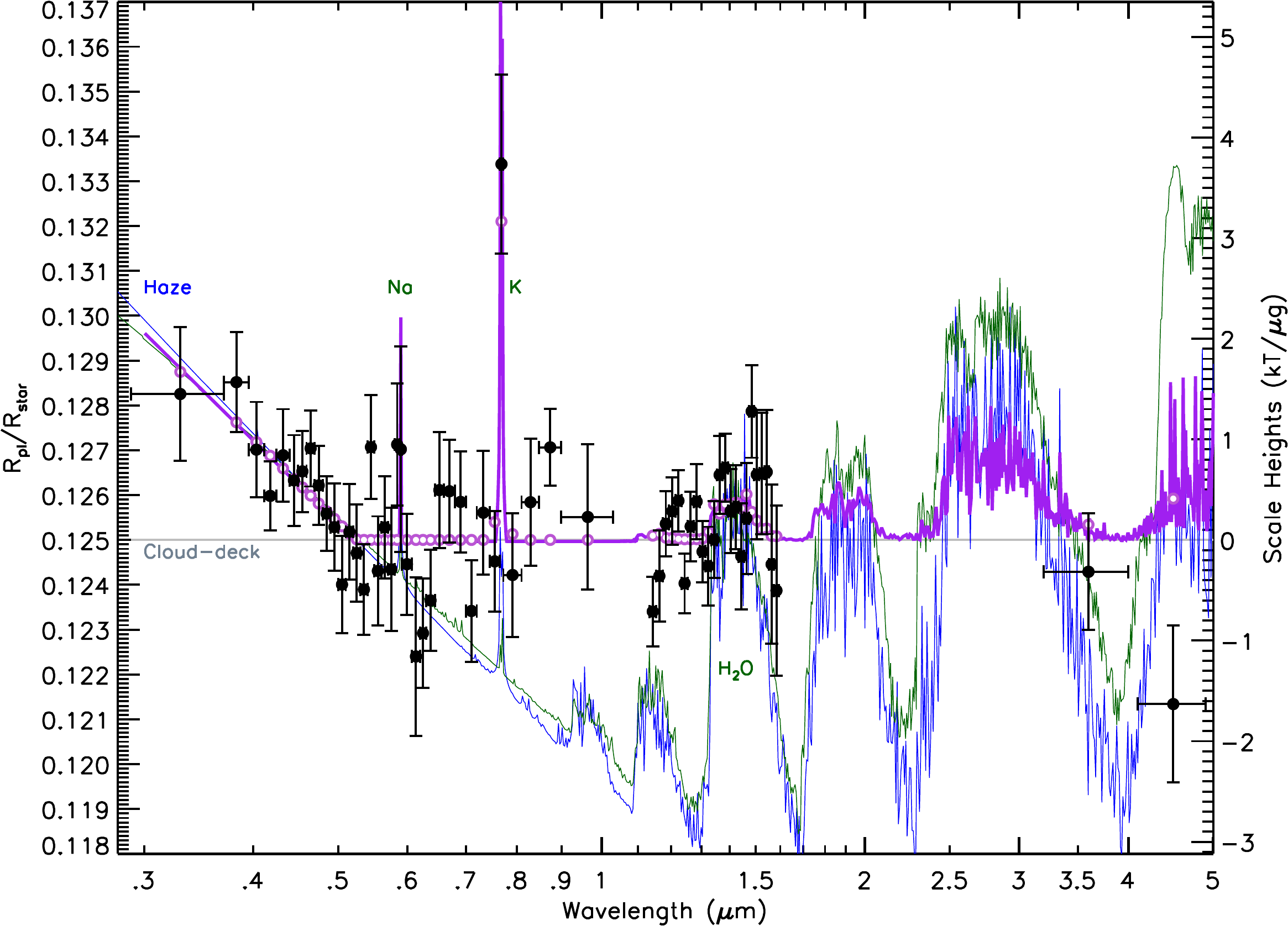}}
\caption[]{Plotted is the broad-band transmission spectral data
  along with atmospheric models.   Two solar-composition
  models containing a scattering haze are shown from the modeling suits of
  Burrows et al. (green) and Fortney et al. (blue).  Our best fit
  model is also plotted (purple) containing a Rayleigh scattering haze, a grey
  cloud-deck at low pressures, non-pressure broadened Na and K
  features, and an obscured H$_2$O feature.  The band-averaged model points are indicated
  with open circles.}
\label{Figure:CH}
\end{figure*}

%%%%%%%%%%%%%%%%%%%%%%%%%%%%%%%%%%%%%%%%%%%%%%%%%%%%%%%%%%%%
\subsection{The Na/K abundance ratio}
Transmission spectra can provide robust measurements of the abundance
ratio between Na and K, which following equation \ref{Eq:Lecav} gives
\begin{equation}
\frac{\varepsilon_{\rm Na}}{\varepsilon_{\rm K}}=
{\frac{\sigma_{\rm K}}{\sigma_{\rm Na}} \exp{\left(
    \frac{z_{\rm Na}-z_{\rm K}}{H} \right)} }, 
\end{equation}
making the ratio dependant on the
relative opacities of the species in question, the measured altitude
difference between the two species, and the scale height of the
atmosphere.  A reliable measurement of the abundance ratio is
particularly useful for WASP-31b, as the source of the Rayleigh
scattering is unknown as is the absolute value of $P_{\rm ref}$ thereby
the absolute abundances are unknown.  However, the relative abundance ratio is
not dependant upon $P_{\rm ref}$ and thereby can still be measured.

To measure the relative abundances, we fit an analytic model which
included Rayleigh scattering, a cloud deck as well as Na and K (see
Fig. \ref{Figure:CH}).  We
also included the small H$_2$O feature from the cloudy Fortney et al. model as found in section
\ref{Sec:cloud} as it provided the best fits, though very similar
results were obtained by removing the opacity of H$_2$O and fitting
the entire infrared wavelength region with a flat line.
Fitting for the alkali abundances, $T$, $\lambda_{\rm T}$
and the baseline radius while setting $P_{\rm ref}$ below 10 mbar gives our best-fitting
overall model to all 61 datapoints, with a $\chi^2$ of 63.0 for 56
d.o.f.  We find a Na to K abundance ratio\footnote{We quote logarithmic abundance values in base e rather than the more common base
  10 given the exponential relationship to the measured values of $z$.}
of ln[Na/K]$=-3.3\pm$2.8 which is
sub-solar compared to the solar value being
ln[Na/K]$_{\rm{\odot}}=+2.8$ \citep{2009ARA&A..47..481A}.  %Asplund.
The uncertainty in the abundance ratio is dominated by the uncertainty
in the Na line core measurement, which could benefit greatly from higher
resolution data clearly separating any Na signatures from the
surrounding clouds.  

With the K line included in the fit, the fit temperature increases
somewhat to $T=2470\pm590$ K.  At the high altitudes probed in the K
line core, the presence of a thermosphere is possible and may be
responsible for the high measured temperatures, as has been
found in HD~189733b and HD~209458b using the Na line \citep{2012MNRAS.422.2477H,2011A&A...527A.110V}.  %Huitson & Vidal
However, the errors on the retrieved temperatures are still high which
limits the current interpretation.  

A sub-solar value for the Na to K abundance ratio is surprising, and to the best of our knowledge has not yet been seen in any planetary
atmosphere.  A variety of physical processes could in principle alter
the abundances, though a Na/K ratio $< 1$ is in any case still
difficult to explain. 
While the alkali metals are easily ionised, which
could dramatically alter their measured neutral abundances,
photo-ionisation favours depleting K~{\small I} over Na~{\small I} presumably making it
a difficult mechanism to explain the observed ratio.
Condensation of Na could also potentially play a role in reducing its
abundance, and would be consistent with the presence of a cloud deck
or higher altitude scatterer.
However, the expectations from current chemical models predict
Na to condense at temperatures near $\sim1000$~K
\citep{1999ApJ...519..793L} % Lodders
which is lower than the expected temperatures of WASP-31b, though a
large day/night temperature contrast could be present, making for a Na
cold-trap on the night side while K is not likely to condense out at
temperatures exceeding $\sim$800~K.  This explanation of condensing Na
but not K is very specific: the planet has to have a specific
temperature regime so that Na$_2$S condenses but not KCL, which
requires only a $\sim$200~K lower temperature.
Finally, we speculate that the large element-to-element
metallicity variations could be primordial in nature, or altered
after formation from accretion processes. 

While the absolute abundances in WASP-31b are unknown, as $P_{\rm ref}$ is unknown,
limits on the pressure can in principle provide bounds on the K abundance.   A cloud-deck pressure level
of 10 mbar or lower, required to hide the pressure-broadened alkali wings, corresponds
to a minimum K abundance of ln[$\varepsilon_{\rm
  K}$]$=-14.77\pm0.86$, indicating a super-solar K abundance (solar ln[$\varepsilon_{\rm
  K}$]$_{\rm{\odot}}=-16.05$).  
The K abundance would be a factor of $\sim$10 higher if the cloud deck
is instead at 1 mbar, required to hide a solar-abundance of H$_2$O.

%%%%%%%%%%%%%%%%%%%%%%%%%%%%%%%%%%%%%%%%%%%%%%%%%%%%%%%%%%%%

\subsection{Cloud properties and vertical mixing}
A broad-band transmission spectrum can give valuable constraints on
aerosol particle sizes.
Obtaining a flat spectrum out to wavelengths of
several microns requires a particle size of at least $\sim$1~$\mu$m;
small particles of, e.g., $\sim$0.1~$\mu$m or less would become transparent
in the IR and thus would be inefficient at scattering the radiation,
as necessary to produce a flat spectrum.
The presence of Rayleigh scattering can also give constraints, as scattering occurs when the particle size is less than the wavelength
of light probed.  In the WASP-31b transmission spectrum, there are two
distinct possible scattering slopes in the data, the near-UV to
blue slope as discussed in Sec. \ref{Sec:RayCl}, and second possible
(though fairly flat) slope in the IR as the 4.5~$\mu$m channel exhibits the smallest
radii in our spectrum (at $\sim$2-$\sigma$ significance).

To further estimate the possible particle sizes allowed in the data, we computed a series of transmission spectra assuming an opacity
dominated by a single size aerosol calculated with Mie theory, and
compared it with the transmission spectrum.  
As the cloud and haze composition can not be easily identified, we
explored a variety of plausible materials, including Al$_2$O$_3$, CaTiO$_3$, Fe$_2$O$_3$, MgSiO$_3$, and
a Titan tholin \citep{2013MNRAS.436.2956S}.  %Sing
We find a single particle size does not give a good quality fit to the broad-band spectrum, and two distinct
particle sizes are needed, a larger particle size for the cloud deck
and a smaller size particle for the higher altitude haze layer.
For the condensate materials explored, we find 
cloud particle sizes between 0.52 and 1~$\mu$m are needed to fit the flat transmission spectrum between
0.52 to 3.6~$\mu$m and the small radius at 4.5~$\mu$m.
Fitting for the data $<$0.52~$\mu$m, we find haze particle sizes of 0.02
to 0.1~$\mu$m can reproduce the near-UV to blue optical scattering
slope, with condensates containing lower values for the imaginary part of
the refraction index (like MgSiO$_3$ or Fe-poor Mg$_2$SiO$_4$ silicates) giving larger particle sizes.

\begin{figure}
\begin{centering}
\includegraphics[width=0.49\textwidth]{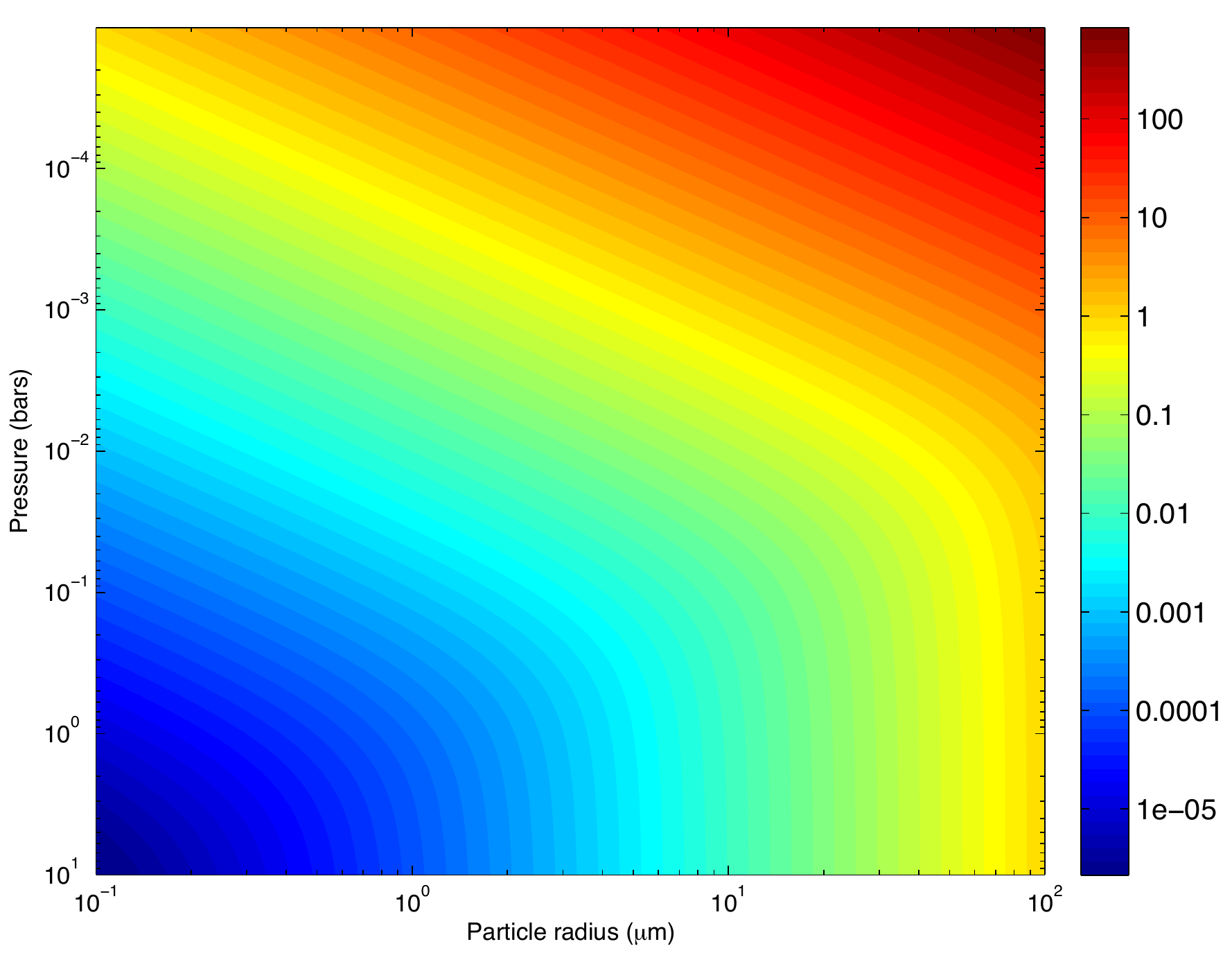}
\par\end{centering}
\caption{Downward settling velocity (colorscale, $\rm m\,s^{-1}$)
for spherical atmospheric particles as a function
of particle radius and pressure, assuming an H$_2$ atmosphere with
a temperature of 1570 K and gravity of $4.56\rm\,m\,s^{-2}$, appropriate
to WASP-31b (see Equations~(3--7) of \protect\citealt{2013A&A...558A..91P}).  %Parmentier et al. (2013).
Stokes flow holds at high pressure and leads to settling velocities that 
depend on particle size but not pressure.  Mean-free-path effects
become important at low pressure, where settling velocity scales as approximately
one over pressure.  Here we have assumed a particle density of 3000~kg~m$^{-3}$ (appropriate to silicate) and the gas viscosity appropriate to
hydrogen from \protect\cite{Rosner2000}; see \protect\cite{2013A&A...558A..91P} %Parmentier et al. (2013) 
for more details.}
\label{fallspeed}
\end{figure}

Our results can also place important constraints on the vertical mixing rate in
the atmosphere of WASP-31b.  As discussed previously, a flat spectrum
requires the presence of cloud or haze particles down to the 10-mbar
level (based on the absence of the potassium line wings) or potentially even
up to the 1-mbar level (based on the weakness of the water spectral
feature).  
Figure~\ref{fallspeed} shows the expected particle settling velocity for
conditions relevant to WASP-31b as a function of particle radius and
pressure, based on Stokes-Cunningham theory.  The regime in the bottom
portion of the plot corresponds to Stokes flow, where settling
velocity is approximately independent of pressure.  The regime at the
top corresponds to that where the mean-free path becomes significant
compared to the particle size, such that the continuum approximation
breaks down and gas-kinetic effects are important; in this regime, the
settling velocity for a particle of a given size scales approximately
with the inverse of pressure.  The figure shows that the settling
velocity of 1-micron particles is 0.01 m/s at 10 mbar and 0.1 m/s at
1 mbar.

Given these estimates of settling velocity, it is interesting to
estimate the effective atmospheric eddy diffusivities that would be
necessary to loft such 1-micron particles to 1-10 mbar pressures, as
necessary to cause a flat transmission spectrum.
\cite{2013A&A...558A..91P} % Parmentier et
considered an idealised 1D diffusion model with downward
transport due to particle settling being balanced by upward transport
due to mixing by atmospheric motions (see Eq. 10 in
\citealt{2013A&A...558A..91P}).  %Parmentier et
An order-of-magnitude consideration of this equation
suggests that, crudely, to keep particles of a given size lofted to
a given altitude, one requires an eddy diffusivity of order
$K_{\rm zz} \sim H w_{\rm settle}$, where $H$ is the atmospheric scale
height and $w_{\rm settle}$ is the particle settling velocity (e.g.,
as shown in Figure~\ref{fallspeed}).  Given the $\sim$1000-km scale height expected on
WASP-31b and the settling velocities quoted above, we estimate that
the necessary atmospheric mixing corresponds to $K_{\rm zz} \sim 10^4
\rm \,m^2\,s^{-1}$ to loft 1-micron particles to 10 mbar and $K_{\rm zz} \sim
10^5 \rm \,m^2\,s^{-1}$ to loft them to 1 mbar.  These estimates
are consistent with the more detailed analysis presented in
\cite{2013A&A...558A..91P}.  %Parmentier

Determining whether the atmospheric circulation on WASP-31b can
generate sufficient mixing to induce $K_{\rm zz}\sim
10^4$--$10^5\rm\,m^2\,s^{-1}$ is not trivial and will require detailed
3D numerical simulations for this planet, which is a task for the
future.  Nevertheless, \cite{2013A&A...558A..91P}  %Parmentier et
made such estimates
for HD 209458b, which has a similar effective temperature to WASP-31b.
Because the incident starlight drives the vertical mixing, we might
expect that planets with similar incident stellar irradiation, such as
WASP-31b and HD 209458b, should exhibit similar profiles of the
vertical mixing rate.  \cite{2013A&A...558A..91P} %Parmentier et
showed that, interestingly, $K_{\rm zz}$ is {\it not} well represented by the
product of the scale height and the rms vertical velocity of the
atmospheric motions (as has sometimes been assumed in the literature).
They fit diffusion models to their 3D simulations of the atmospheric
circulation to obtain the vertical profile of the effective
diffusivity: $K_{\rm zz} \approx (5\times10^4/\sqrt{P_{\rm bar}})
\rm\,m^2\,s^{-1}$, where $P_{\rm bar}$ is the pressure in bar.  This
equation suggests that the effective eddy diffusivity should be
$\sim$$5\times10^5\rm\,m^2\,s^{-1}$ at 10 mbar and
$\sim$$2\times10^6\rm\,m^2\,s^{-1}$ at 1 mbar.  These values exceed
the estimates we made above for the diffusivity required to mix
$\sim$1-micron particles to pressures of 1--10 mbar.  Thus, it appears
physically plausible that the circulation will indeed be sufficiently
strong to mix particles upward to pressures of 1--10 mbar, as necessary
to explain our flattened spectrum.

%%%%%%%%%%%%%%%%%%%%%%%%%%%%%%%%%%%%%%%%%%%%%%%%%%%%%%%%%%%%

\section{CONCLUSION}
We have presented {\it HST} and Spitzer observations totalling six transits of
WASP-31b which cover the full-optical to infrared spectral regions.  The
resulting exoplanet transmission spectrum is unique, showing several atmospheric features
not yet seen in conjunction including potassium line cores and no
detectable sodium, a flat optical to near-IR slope indicative of a cloud deck, and Rayleigh scattering at
short wavelengths.
The potassium feature is strong (detected at 4.2-$\sigma$ confidence)
though lacks significant pressure broadened wings, which we find limits the
pressures probed to 10-mbar or less.
Weak sodium and strong potassium lines indicate the
peculiar possibility of a sub-solar [Na/K] abundance ratio below one, which is
difficult to explain though may point toward Na condensation or
large element-to-element metallically variations from
formation/accretion processes.   

The transmission spectrum is flat longward of 0.52$\mu$m, indicative
of a high altitude broad-band cloud deck which covers the expected optical Na and
near-IR H$_2$O features.  However, despite containing a substantial cloud(s) the broad-band transmission
spectrum is not flat, and a 4.2-$\sigma$ significant slope is observed
at wavelengths shortward of 0.52$\mu$m, consistent with Rayleigh
scattering by small sub-micron size haze particles.

WASP-31b joins a growing group of hot-Jupiter exoplanets with evidence
for clouds and hazes.  Clouds and hazes are now seen to be important
atmospheric features in hot Jupiters across a wide range of
effective temperatures, including the hottest planets like WASP-12b
and much cooler planets like HD~189733b.  Moreover, at temperatures
near 1570~K, WASP-31b has an equilibrium temperature close to the
mean for known hot Jupiters, implying atmospheric clouds and haze
signatures are not confined to the extremes of hot Jupiter parameter space.  

We highlight the importance of wide wavelength coverage for exoplanet transmission spectra, as
short-wavelength data can be highly sensitive to the cloud properties and the blue-optical data was necessary in this case to
detect aerosol scattering features.

\section*{ACKNOWLEDGEMENTS}

This work is based on observations
with the NASA/ESA HST, obtained at the
Space Telescope Science Institute (STScI) operated by AURA, Inc.  
This work is also based in part on observations made with the Spitzer Space Telescope, which is operated by the Jet Propulsion Laboratory, California Institute of Technology under a contract with NASA.
The research leading to these results has received funding from the European Research Council under the European Union's Seventh Framework Programme (FP7/2007-2013) / ERC grant agreement n$^{\circ}$ 336792.
We thank the anonymous referee for their comments.
DS, FP and, NN acknowledge support from STFC consolidated grant ST/J0016/1.
Support for this work was provided by NASA through grants under the HST-GO-12473 programme from
the STScI.  
PW and HW acknowledge support from the UK Science and Technology Facilities Council (STFC). 
ALE and AVM acknowledge support from the French Agence Nationale de la Recherche (ANR), under programme ANR-12-BS05-0012 `Exo-Atmos'.

\footnotesize{
\bibliographystyle{yahapj} % style citations
\bibliography{Wasp31.manuscript.V6} % your references .bib
}

\begin{table*} %==================================================
\caption{WASP-31b broad-band transmission spectral results and non-linear limb darkening
  coefficients for the STIS $G$430$L$, $G$750$L$, WFC3 $G$141, and Spitzer IRAC.}
\label{Table:LDTrans}
\begin{centering}
\renewcommand{\footnoterule}{}  % to avoid a line before footnotes
\begin{tabular}{ccrrrr}
\hline\hline  %==================================================
 $\lambda$(\AA) & $R_{P}/R_{*}$  &$c_1$ & $c_2$ & $c_3$ & $c_4$ \\
\hline  %==================================================
%\hline%======================%common mode ===========
 2900 -  3700    & 0.12825 $\pm$ 0.00149 & 0.3403& 0.6641&-0.0625&-0.0907\\
 3700 -  3950    & 0.12852 $\pm$ 0.00111 & 0.3957&-0.0018& 0.9565&-0.4757\\
 3950 -  4113    & 0.12701 $\pm$ 0.00106 & 0.2129& 0.7661&-0.0385&-0.0737\\
 4113 -  4250    & 0.12598 $\pm$ 0.00077 & 0.2136& 0.8688&-0.1927&-0.0317\\
 4250 -  4400    & 0.12689 $\pm$ 0.00103 & 0.4479&-0.2031& 1.1499&-0.5530\\
 4400 -  4500    & 0.12633 $\pm$ 0.00101 & 0.2837& 0.6062& 0.0919&-0.1409\\
 4500 -  4600    & 0.12653 $\pm$ 0.00091 & 0.1778& 1.1859&-0.7235& 0.1887\\
 4600 -  4700    & 0.12704 $\pm$ 0.00085 & 0.2336& 1.0124&-0.5540& 0.1291\\
 4700 -  4800    & 0.12622 $\pm$ 0.00088 & 0.2531& 0.9945&-0.5675& 0.1259\\
 4800 -  4900    & 0.12559 $\pm$ 0.00083 & 0.2632& 0.9362&-0.4808& 0.0835\\
 4900 -  5000    & 0.12529 $\pm$ 0.00097 & 0.3486& 0.7057&-0.2924& 0.0280\\
 5000 -  5100    & 0.12401 $\pm$ 0.00108 & 0.3328& 0.7698&-0.3730& 0.0554\\
 5100 -  5200    & 0.12518 $\pm$ 0.00107 & 0.4177& 0.4396& 0.0065&-0.0924\\
 5200 -  5300    & 0.12471 $\pm$ 0.00109 & 0.3246& 0.8223&-0.4701& 0.0936\\
 5300 -  5400    & 0.12390 $\pm$ 0.00101 & 0.3897& 0.5857&-0.2007&-0.0098\\
 5400 -  5500    & 0.12707 $\pm$ 0.00116 & 0.3873& 0.5986&-0.2214&-0.0062\\
 5500 -  5600    & 0.12431 $\pm$ 0.00122 & 0.3640& 0.6822&-0.3215& 0.0315\\
 5600 -  5700    & 0.12528 $\pm$ 0.00114 & 0.3673& 0.7158&-0.4132& 0.0757\\
 5700 -  5800    & 0.12434 $\pm$ 0.00137 & 0.4070& 0.5534&-0.2195& 0.0001\\
 5800 -  5878    & 0.12713 $\pm$ 0.00137 & 0.3774& 0.6655&-0.3623& 0.0542\\
 5878 -  5913    & 0.12702 $\pm$ 0.00229 & 0.4276& 0.4659&-0.1253&-0.0414\\%Na
 5913 -  6070    & 0.12446 $\pm$ 0.00113 & 0.3873& 0.6505&-0.3927& 0.0743\\
 6070 -  6200    & 0.12240 $\pm$ 0.00177 & 0.4058& 0.5806&-0.3146& 0.0386\\
 6200 -  6300    & 0.12292 $\pm$ 0.00122 & 0.4304& 0.4956&-0.2242& 0.0027\\
 6300 -  6450    & 0.12365 $\pm$ 0.00113 & 0.4582& 0.3866&-0.1029&-0.0503\\
 6450 -  6600    & 0.12611 $\pm$ 0.00129 & 0.4111& 0.5841&-0.3151& 0.0084\\
 6600 -  6800    & 0.12609 $\pm$ 0.00114 & 0.4590& 0.3784&-0.1053&-0.0506\\
 6800 -  7000    & 0.12585 $\pm$ 0.00113 & 0.4657& 0.3399&-0.0804&-0.0525\\
 7000 -  7200    & 0.12342 $\pm$ 0.00114 & 0.4961& 0.2114& 0.0595&-0.1054\\
 7200 -  7450    & 0.12561 $\pm$ 0.00138 & 0.4941& 0.2119& 0.0391&-0.1013\\
 7450 -  7645    & 0.12452 $\pm$ 0.00112 & 0.4953& 0.2053& 0.0076&-0.0779\\
 7645 -  7720    & 0.13338 $\pm$ 0.00200 & 0.5184& 0.1285& 0.0841&-0.1041\\%K
 7720 -  8100    & 0.12422 $\pm$ 0.00138 & 0.5115& 0.1318& 0.0783&-0.1059\\
 8100 -  8485    & 0.12584 $\pm$ 0.00142 & 0.5043& 0.1541& 0.0361&-0.0841\\
 8485 -  8985    & 0.12706 $\pm$ 0.00086 & 0.5166& 0.0389& 0.1599&-0.1358\\
 8985 - 10300    & 0.12551 $\pm$ 0.00162 & 0.5165& 0.0417& 0.1293&-0.1206\\
11340 - 11540    & 0.12340 $\pm$ 0.00078 & 0.4324& 0.3175&-0.3283& 0.0956\\
11540 - 11740    & 0.12420 $\pm$ 0.00102 & 0.3932& 0.4846&-0.5483& 0.1904\\
11740 - 11940    & 0.12536 $\pm$ 0.00073 & 0.4458& 0.2965&-0.3251& 0.0941\\
11940 - 12140    & 0.12564 $\pm$ 0.00075 & 0.4645& 0.2109&-0.2184& 0.0507\\
12140 - 12340    & 0.12587 $\pm$ 0.00068 & 0.4574& 0.2195&-0.2376& 0.0664\\
12340 - 12540    & 0.12403 $\pm$ 0.00066 & 0.4180& 0.4200&-0.5092& 0.1784\\
12540 - 12740    & 0.12530 $\pm$ 0.00077 & 0.4263& 0.4442&-0.5611& 0.1938\\
12740 - 12940    & 0.12586 $\pm$ 0.00084 & 0.4589& 0.3528&-0.5107& 0.1923\\
12940 - 13140    & 0.12474 $\pm$ 0.00068 & 0.4270& 0.4563&-0.6173& 0.2321\\
13140 - 13340    & 0.12442 $\pm$ 0.00087 & 0.4253& 0.4880&-0.6681& 0.2545\\
13340 - 13540    & 0.12501 $\pm$ 0.00086 & 0.6197&-0.4348& 0.4174&-0.1567\\
13540 - 13740    & 0.12645 $\pm$ 0.00087 & 0.4314& 0.5228&-0.7475& 0.2933\\
13740 - 13940    & 0.12661 $\pm$ 0.00076 & 0.4524& 0.4801&-0.7299& 0.2940\\
13940 - 14140    & 0.12565 $\pm$ 0.00093 & 0.4788& 0.4154&-0.6779& 0.2773\\
14140 - 14340    & 0.12573 $\pm$ 0.00093 & 0.4815& 0.4623&-0.7680& 0.3219\\
14340 - 14540    & 0.12463 $\pm$ 0.00119 & 0.4997& 0.4165&-0.7444& 0.3207\\
14540 - 14740    & 0.12548 $\pm$ 0.00124 & 0.5228& 0.2094&-0.4124& 0.1611\\
14740 - 14940    & 0.12787 $\pm$ 0.00103 & 0.5299& 0.3487&-0.6754& 0.2930\\
14940 - 15140    & 0.12646 $\pm$ 0.00144 & 0.5531& 0.3286&-0.6973& 0.3124\\
15140 - 15340    & 0.12648 $\pm$ 0.00136 & 0.6050& 0.1936&-0.5897& 0.2824\\
15340 - 15540    & 0.12651 $\pm$ 0.00138 & 0.6699& 0.3129&-0.9263& 0.4493\\
15540 - 15740    & 0.12446 $\pm$ 0.00177 & 0.6603& 0.0663&-0.5009& 0.2625\\
15740 - 15940    & 0.12387 $\pm$ 0.00190 & 0.6583& 0.0797&-0.5063& 0.2577\\
15940 - 16140    & 0.12241 $\pm$ 0.00172 & 0.6465& 0.1055&-0.5546& 0.2837\\
16140 - 16340    & 0.12447 $\pm$ 0.00190 & 0.7357&-0.0020&-0.4485&0.2387\\
36000                  & 0.12429 $\pm$ 0.00131 & 0.3632&-0.1314&  0.0949& -0.0435\\
45000                  & 0.12134 $\pm$ 0.00175 & 0.4514&-0.5278&  0.5414& -0.2073\\
\hline%=========================================================
\end{tabular}
\end{centering}
\end{table*}

\end{document}